\documentclass[aip,preprint,pop]{revtex4-2}
\usepackage{graphicx}
\usepackage{amsmath,amssymb,amsfonts,bm}

\usepackage{caption}
\usepackage[subrefformat=parens]{subcaption}

\newcommand{\pd}{\partial}
\newcommand{\td}{{\rm d}}
\newcommand{\nab}{\nabla}
\newcommand{\pr}{\prime}
\newcommand{\rmi}{{\rm i}\,}
\newcommand{\rme}{{\rm e}}
\newcommand{\vphi}{\varphi}
\newcommand{\veps}{\varepsilon}
\newcommand{\muo}{\mu_{0}}
\newcommand{\vB}{{\bm B}}
\newcommand{\vu}{{\bm u}}
\newcommand{\vv}{{\bm v}}
\newcommand{\vx}{{\bm x}}
\newcommand{\hrvz}{\hat{\bm z}}
\newcommand{\dis}{\displaystyle}

\begin{document}


\title{
Linear stability analysis via simulated annealing and accelerated relaxation
}

\author{M. Furukawa}
\email[]{furukawa@tottori-u.ac.jp}
\affiliation{
Faculty of Engineering, Tottori Univ., 
Minami 4-101, Koyama-cho, Tottori-shi, Tottori 680-8552, Japan
}

\author{P. J. Morrison}
\affiliation{
Department of Physics and Institute for Fusion Studies, Univ. Texas at Austin, 
TX, 78712, USA
}

\date{\today}

\begin{abstract}
Simulated annealing (SA) is a kind of relaxation method for finding equilibria of Hamiltonian systems.  A set of evolution equations is solved with SA, which is derived from the original Hamiltonian system so that the energy of the system changes monotonically while preserving Casimir invariants inherent to noncanonical Hamiltonian systems.  The energy extremum reached by SA is an equilibrium. Since  SA searches for an energy extremum, it can also be used for stability analysis when initiated from a state where a perturbation is added to an equilibrium.  The procedure of the stability analysis is explained, and some examples are shown. Because the time evolution is computationally time consuming,  efficient relaxation is necessary for SA to be practically useful.  An acceleration method is developed  by introducing time dependence  in the  symmetric kernel used in the double bracket, which is part of the SA formulation described here.  An explicit formulation for  low-beta reduced  magnetohydrodynamics (MHD) in cylindrical geometry is presented. Since SA for low-beta reduced MHD has two advection fields that relax, it is important to balance the orders of magnitude of these advection fields.
\end{abstract}


\maketitle

\section{Introduction}
\label{sec:introduction}

Simulated annealing (SA)\cite{Flierl-Morrison-2011} is a kind of relaxation method for obtaining equilibria, stationary states,  of Hamiltonian systems.  In noncanonical Hamiltonian systems such as non-dissipative fluid dynamics (see Ref.~\onlinecite{Morrison-1998} for review), there exists Casimir invariants that foliate the infinite-dimensional phase space.
The surface in the phase space on which each Casimir invariant takes a same value is called a Casimir leaf.  A stationary state or an equilibrium of the system is given by an energy extremum on the Casimir leaf.\cite{KO-1958, Arnold-1965-1}

With SA, a set of artificial evolution equations is solved, which is derived from the original equations of the Hamiltonian system. The set of artificial evolution equations is constructed so that the energy of the system changes monotonically, while preserving the Casimir invariants.  The idea of the artificial dynamics was proposed originally for
obtaining stationary states of two-dimensional fluid flows\cite{Vallis-1989, Carnevale-1990, Shepherd-1990} and generalized in Ref.~\onlinecite{Flierl-Morrison-2011}.  In Ref.~\onlinecite{Flierl-Morrison-2011},  the formulation of the artificial dynamics was considerably extended and made more workable by providing a means of including additional constraints and by introducing a smoothing kernel.  There a variety of stationary states were obtained for two-dimensional
vortex dynamics and associated two-layer problems.  The terminology ``simulated annealing'' was introduced in this paper.

The noncanonical Hamiltonian structure of ideal magnetohydrodynamics
(MHD) was first given in Ref.~\onlinecite{Morrison-1980}.
For testing the idea of SA, the MHD Hamiltonian structure was first used \cite{Chikasue-2015-PoP, Chikasue-2015-JFM} for   low-beta reduced MHD\cite{Strauss-1976} in a two-dimensional rectangular domain with doubly periodic boundary conditions.  Since  low-beta reduced MHD has two fields, i.e., vorticity and magnetic flux function, the relaxation path to a stationary state becomes rather complex.  However, the application was successful. Later it was extended to cylindrical  geometry, and an equilibrium with magnetic islands was obtained.\cite{Furukawa-2017} Further,  SA was applied to  high-beta reduced MHD\cite{Strauss-1977} in toroidal geometry, and successfully used to obtain tokamak as well as toroidally averaged stellarator equilibria.\cite{Furukawa-2018}

The above results  are based on the double bracket formulation of SA.\cite{Morrison-2017} However, another type of artificial dynamics method for calculating stationary states is  metriplectic dynamics.\cite{Morrison-1986-2,Morrison-2017}
This relaxation method was successfully applied to two-dimensional vortex dynamics and further to axisymmetric MHD equilibria\cite{Bressan-2018} described by the Grad-Shafranov equation.\cite{Shafranov-1958, Lust-1957, Grad-1958}

From a more general point of view,
minimization of a functional like the energy functional for SA in this paper,
may be done by various methods
such as a classical steepest-descent method and its extensions,
the Nelder-Mead method\cite{Nelder-Mead-1965},
genetic algorithms\cite{Holland-1975}
and many others developed in the context of machine learning.
Although they can be used for minimizing the energy functional of the
MHD systems, it may not be straightforward to preserve the Casimir
invariants since these methods are not based on the Hamiltonian structure of the
governing equations.  Our method is a natural choice for the energy
minimization on a Casimir leaf.

It is pointed out that because SA searches for an energy extremum, or an
energy minimum in the MHD problems, it can be used not only for
equilibrium calculations but also for stability analysis.  Suppose an
equilibrium is given by solving the Grad-Shafranov equation in toroidal
geometry or, e.g., just as a trivial solution in cylindrical geometry.
Then, suppose a perturbation is given to the equilibrium followed by the
SA procedure.  If the equilibrium is an energy minimum,  SA will recover
the original equilibrium.  If not,  SA will lead to another state that
is an energy minimum, or at least will leave the original equilibrium in
a sense of linear stability.  Note that it is
important to stay on the same Casimir leaf as the original equilibrium when giving the perturbation. Such perturbations were termed dynamically accessible in Refs.~\onlinecite{MP89,MP90,AMP16}.  If a perturbation is not dynamically accessible,  SA will not  recover the original equilibrium because the perturbed state is on another Casimir leaf.  The SA procedure will find another equilibrium that is on the Casimir leaf where the perturbed state exists. In the present paper, the procedure for the stability analysis, especially how the perturbation is given without leaving the Casimir leaf, is developed.

On the basis of the previous studies, one would expect that SA can calculate a wide class  of ideal MHD equilibria including magnetic islands and/or magnetic chaotic regions.  Also,  SA can be used for analysis of not only linear but also nonlinear stability.   However, it is not so straightforward because  SA solves an initial-value problem, and the numerical simulation
of an initial-value problem generally takes time.
Even in the linear stability analysis, it takes time to show that an
equilibrium is stable since we have to observe that the given
perturbation disappears.
Therefore, an acceleration method is necessary for  SA to be
practically useful.  In the present paper, an acceleration method is developed for the double bracket formulation of SA.
The method is explicitly described for the low-beta reduced MHD model in cylindrical geometry, and applied to the stability analysis.

The paper is organized as follows.  In Sec.~\ref{sec:model}, the double
bracket formulation of SA for low-beta reduced MHD  is introduced.  Then
the procedure for the stability analysis is explained in
Sec.~\ref{sec:procedure}. Section~\ref{sec:results} presents numerical
results.
Linear stability of a stable equilibrium is analyzed in 
Sec.~\ref{subsec:stableCase}, while
Sec.~\ref{subsec:unstableCase} is for an unstable equilibrium.
An equilibrium, the stability of which is to be subsequently analyzed,  is given in Sec.~\ref{subsubsec:equilibrium}, and a  dynamically accessible perturbation is given to the equilibrium in
Sec.~\ref{subsubsec:perturbation}.  Here it is explained how the dynamically accessible perturbation  stays on the  original Casimir leaf.   Then, a preliminary result, one before an acceleration method is introduced, is presented,
and the problem of slow convergence is explicitly shown in Sec.~\ref{subsubsec:slowConvergence}.  The slow convergence comes from an imbalance of relaxation speeds between the relaxing fields. In order to confirm this, simulation results where one  field  is eliminated is presented in Sec.~\ref{subsubsec:forbiddenDirection}. Next, in Sec.~\ref{subsubsec:acceleration},
the acceleration method is developed that balances the relaxation
speeds, and numerical results are shown.
Section~\ref{subsubsec:anotherInitialPerturbation} shows numerical results
with another initial condition where the kinetic energy perturbation is
larger. 
As explained earlier, Sec.~\ref{subsec:unstableCase} shows that a given
perturbation grows while the energy decreases by SA for an unstable equilibrium.
Stability analyses via SA in this paper are examined in Section~\ref{subsec:stability}.
Section~\ref{sec:conclusion} concludes the paper.

\section{Model}
\label{sec:model}

\subsection{Low-beta reduced MHD and normalization}
\label{subsec:rMHD}

In this paper, the low-beta reduced MHD model\cite{Strauss-1976} is investigated.   We consider a cylindrical plasma with minor radius $a$ and length $2\pi R_{0}$.   The low-beta reduced MHD equations are
\begin{align}
 \frac{\pd U}{\pd t}
&=
 [ U , \vphi ] + [ \psi , J ] - \veps \frac{\pd J}{\pd \zeta},
\label{eq:vorticity-equation}
\\
 \frac{\pd \psi}{\pd t}
&=
 [ \psi , \vphi ] - \veps \frac{\pd \vphi}{\pd \zeta},
\label{eq:Ohm-law}
\end{align}
where all physical quantities are normalized by using the length $a$, the magnetic field in the axial direction $B_{0}$, the Alfv\'en velocity  $v_{\rm A} :=  {B_{0}}/{\sqrt{ \muo \rho_{0} }}$  with $\muo$ and $\rho_{0}$ being vacuum permeability and typical mass density, respectively, and the Alfv\'en time  $\tau_{\rm A} := {a}/{v_{\rm A}}$.  The cylindrical coordinates 
$(r, \theta, z)$ as well as the toroidal angle $\zeta := {z}/{R_{0}}$ are used.
The inverse aspect ratio is defined as  $\veps := {a}/{R_{0}}$.   The fluid velocity is $\vv = \hrvz \times \nab \vphi$, 
and the magnetic field is $\vB = \hrvz + \nab \psi \times \hrvz$, where the unit vector in the $z$ direction is denoted by $\hrvz$. The vorticity is $U := \bigtriangleup_{\perp} \vphi$ and   the current density is $J := \bigtriangleup_{\perp} \psi$, where the Laplacian in the $r$--$\theta$ plane is denoted by $\bigtriangleup_{\perp}$. The Poisson bracket for two functions $f$ and $g$ is defined as  $[ f, g ] := \hrvz \cdot \nab f \times \nab g$.

The Hamiltonian structure for low-beta reduced MHD, which follows from
that of Ref.~\onlinecite{Morrison-1980}, was  first given in
Refs.~\onlinecite{MH84,MM84}.  The  Hamiltonian is the summation of kinetic and magnetic energies as
\begin{equation}
 H[ \vu ] 
:= 
 \int_{\cal D} \td^{3}x \,
      \frac{1}{2} \left\{
         \left| \nab_{\perp} ( \bigtriangleup_{\perp}^{-1} U ) \right|^{2}
       + \left| \nab \psi \right|^{2}
                  \right\},
\label{eq:Hamiltonian}
\end{equation}
where ${\cal D}$ is the domain of the cylindrical plasma.
The Casimir invariants are
\begin{align}
 C_{\rm m} &:= \int_{\cal D} \td^{3}x \, \psi(\vx, t),
 \\
 C_{\rm v} &:= \int_{\cal D} \td^{3}x \, U(\vx, t).
\end{align}
Furthermore,
\begin{align}
 C_{1} &:= \int_{\cal D} \td^{3}x \, f(\psi_{\rm h}),
 \\
 C_{2} &:= \int_{\cal D} \td^{3}x \, U g(\psi_{\rm h})
\end{align}
are also Casimir invariants 
when only single helicity components are included in the dynamics.
Here $f$ and $g$ are arbitrary functions of a helical flux 
$\psi_{\rm h}(\vx, t) := \psi(\vx, t) + \veps r^{2} / (2 q_{\rm s})$ for
a specified 
safety factor of a rational number $q_{\rm s} = m/n$ with $m$ and $n$
being a poloidal and a toroidal mode number, respectively.

\subsection{Double bracket formulation of simulated annealing}
\label{subsec:SA}

We adopt simulated annealing,  formulated by using a double bracket\cite{Flierl-Morrison-2011, Morrison-2017, Furukawa-2017}.  For  low-beta reduced MHD, the set of artificial evolution equations are conveniently given in the reduced  following form:
\begin{align}
 \frac{\pd U}{\pd t}                                                           
 &=
 [ U, \tilde{\vphi} ] + [ \psi, \tilde{J} ]
 - \veps \frac{\pd \tilde{J}}{\pd \zeta},
\label{eq:SA-vorticity-equation}
 \\
 \frac{\pd \psi}{\pd t}
 &=
 [ \psi, \tilde{\vphi} ] - \veps \frac{\pd \tilde{\vphi}}{\pd \zeta},
\label{eq:SA-Ohm-law}
\end{align}
where the advection fields are defined as
\begin{align}
 \tilde{\vphi}(\vx, t)
 &:=
 \int_{\cal D} \td^{3} x^{\pr} \,
 K_{1j}(\vx, \vx^{\pr}) f^{j}(\vx^{\pr}, t),
\label{eq:tvphi}
 \\
 \tilde{J}(\vx, t)
 &:=
 \int_{\cal D} \td^{3} x^{\pr} \,
 K_{2j}(\vx, \vx^{\pr}) f^{j}(\vx^{\pr}, t).
\label{eq:tJ}
\end{align}
Here $j=1$ or $2$, $f^{1}$ is the right-hand side of Eq.~(\ref{eq:vorticity-equation}), $f^{2}$ is the right-hand side of Eq.~(\ref{eq:Ohm-law}), and $(K_{ij})$ is a kernel with a definite sign.

Note that the equations have the same form as those of the original low-beta reduced MHD; however,  the advection fields are replaced by the artificial ones. The advection fields must be chosen so that the energy of the system changes monotonically, however, there are still a variety of choices through the kernel.

If $(K_{ij})$ is positive definite, the energy of the system decreases monotonically.  On the other hand, Casimir invariants are preserved. In this study, we choose $(K_{ij})$ to be diagonal with its diagonal components given by 
\begin{equation}
 K_{ii}(\vx, \vx^{\pr}) = \alpha_{ii} g(\vx, \vx^{\pr})\,,
\end{equation}
for $i=1$ and $2$, where $g(\vx, \vx^{\pr})$ is a Green's function defined through 
\begin{equation}
 \bigtriangleup g(\vx, \vx^{\pr})
  := -\delta^{3}(\vx - \vx^{\pr}).
\end{equation}
Here, $\bigtriangleup$ is the Laplacian and  $\delta^{3}(\vx)$ is the Dirac's delta function in three dimensions, respectively.

\section{Stability analysis procedure}
\label{sec:procedure}

In this section, the procedure for using SA to assess stability is  explained.  It consists of three steps. 

The first step is to choose an equilibrium. Although SA has been used for obtaining equilibria; here  the equilibrium can be given by any method. For example, any cylindrically symmetric state is a trivial equilibrium in a cylindrical plasma.  Although unbeknownst beforehand, the chosen equilibrium can be either stable or unstable.  Similarly, any solution to the Grad-Shafranov equation is a candidate equilibrium in axisymmetric toroidal plasma, one for which the SA stability method could be applied. Of course, we could also choose an equilibrium that is obtained by SA.

The second step is to perturb the equilibrium. The key point is to make the perturbation dynamically accessible, i.e., have it stay on the same Casimir leaf as the equilibrium whose stability is being analyzed. This is accomplished by time evolution of $U$ and $\psi$ under Eqs.~(\ref{eq:SA-vorticity-equation}) and (\ref{eq:SA-Ohm-law}) using arbitrary advection fields $\tilde{\vphi}$ and $\tilde{J}$ that are different from Eqs.~(\ref{eq:tvphi}) and (\ref{eq:tJ}). This is possible because the Casimir invariants are preserved for any time evolution by the equations of the form (\ref{eq:SA-vorticity-equation}) and (\ref{eq:SA-Ohm-law}) since they can be generated by the noncanonical Poisson bracket operator, the co-kernel of which projects onto the symplectic leaves.  If we choose these advection fields in the form (\ref{eq:tvphi}) and (\ref{eq:tJ}) with the kernel $(K_{ij})$ of a definite sign, then the energy changes monotonically and the Casimir invariants are preserved.  For the arbitrary choices of $\tilde{\vphi}$ and $\tilde{J}$, the Casimir invariants are still preserved, although  the energy  need not change  monotonically.

The final step is to perform SA. If the chosen equilibrium is an energy minimum state, which by Dirichlet's theorem is a stable equilibrium (cf. Ref.~\onlinecite{Morrison-1998}),   SA starting from the perturbed state in the second step will recover the original minimum energy state. This is at least true when the perturbation is in a linear regime.

Note that SA can identify a linearly unstable case without a big
computational cost 
since a given perturbation grows in time while the energy is decreased
by SA.  
On the other hand, it is time consuming for a stable case since
we have to observe that the given perturbation really disappears by SA.
Even if the perturbation becomes smaller by the SA time evolution,
the system cannot be said to be stable if the perturbation remains with a
small amplitude.

Let us emphasize the importance of  the perturbation being dynamically accessible, where the time evolution in the final step occurs on a Casimir leaf determined by the  initial condition used for  the  SA procedure.  If the perturbed state  were on a different Casimir leaf,  the time evolution of SA  in the final step would not recover the original equilibrium.

\section{Numerical results}
\label{sec:results}

\subsection{A case of stable equilibrium}
\label{subsec:stableCase}

\subsubsection{A case study equilibrium for stability analysis}
\label{subsubsec:equilibrium}

\begin{figure}
 \centering
 \includegraphics[width=0.45\textwidth]{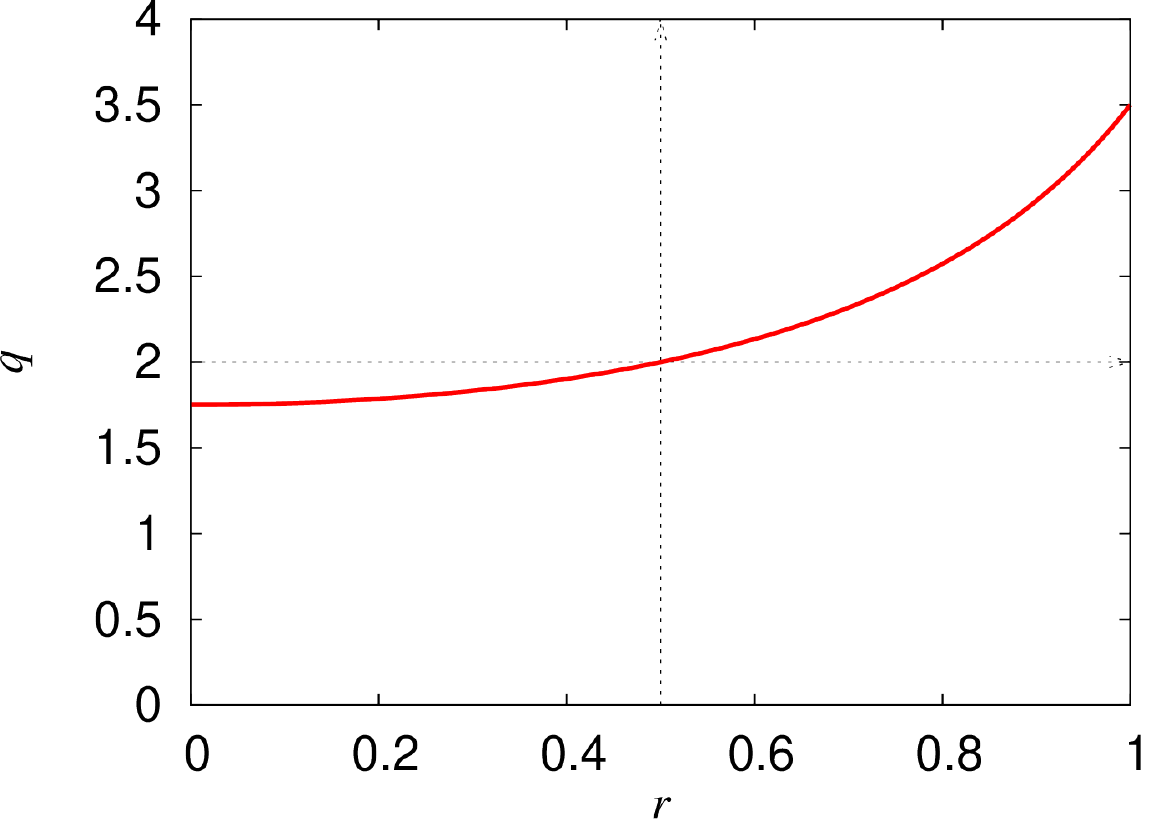}
 \caption{Safety factor $q$ profile. }
 \label{fig:r-q}
\end{figure}

In order to clearly explain the efficacy of our method, in the present paper we choose an equilibrium with cylindrical symmetry.  Any cylindrically symmetric state is an equilibrium of Eqs.~(\ref{eq:vorticity-equation}) and (\ref{eq:Ohm-law}). We assume that the equilibrium has no plasma rotation. The safety factor profile is assumed to be $q(r) = q_{0} / ( 1 - r^{2}/2 )$ with $q_{0} = 1.75$, and shown in Fig.~\ref{fig:r-q}. It has $q=2$ surface at $r=1/2$.

This equilibrium is spectrally stable against ideal MHD modes, although it is unstable to a resonant $(m,n)=(-2,1)$ tearing mode. Note that we take the perturbed quantities proportional to $\rme^{\rmi (m \theta + n \zeta)}$. Since  SA does not change magnetic field line topology,  we expect that  SA will recover this equilibrium as an energy minimum state.

\subsubsection{Dynamically accessible perturbation of the equilibrium}
\label{subsubsec:perturbation}

\begin{figure*}
\begin{minipage}[t]{0.45\textwidth}
 \centering
 \includegraphics[width=\textwidth]{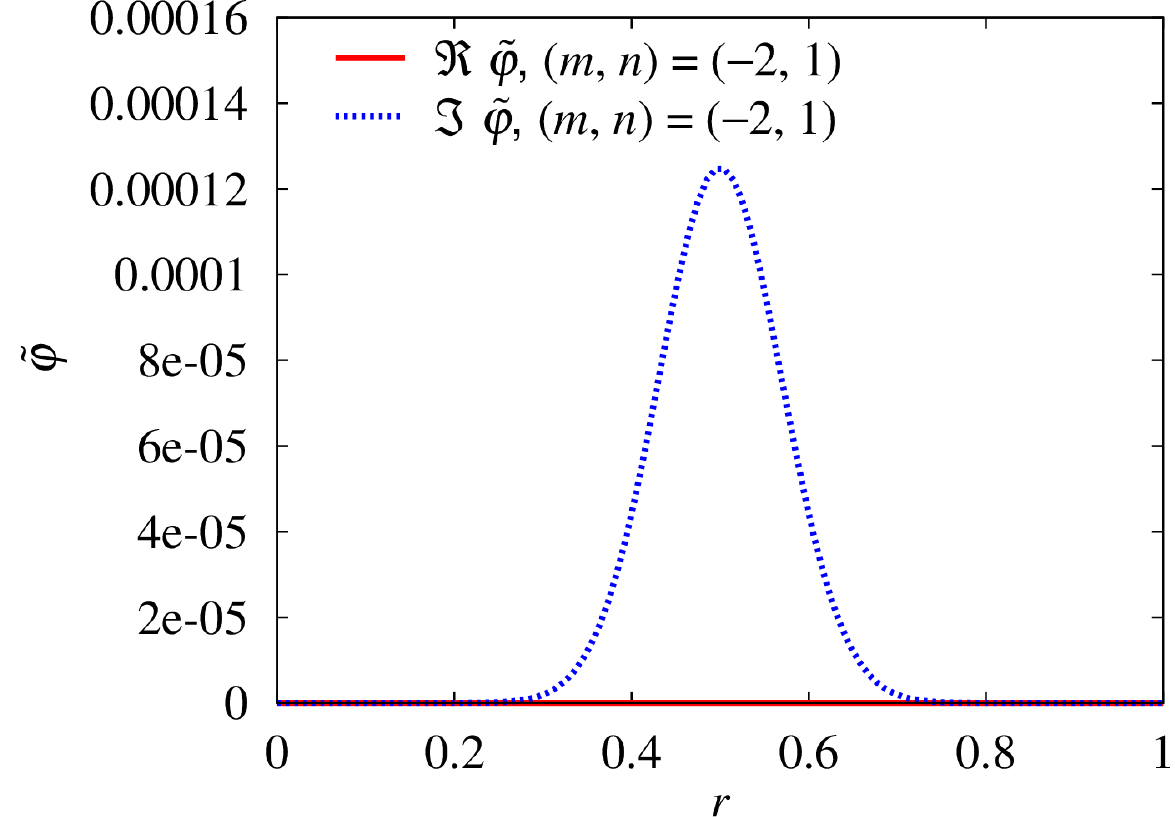}
 \subcaption{$\tilde{\vphi}$. Real part is zero.}
 \label{fig:r-tvphi-perturb}
\end{minipage}
 \hfill
\begin{minipage}[t]{0.45\textwidth}
 \centering
 \includegraphics[width=\textwidth]{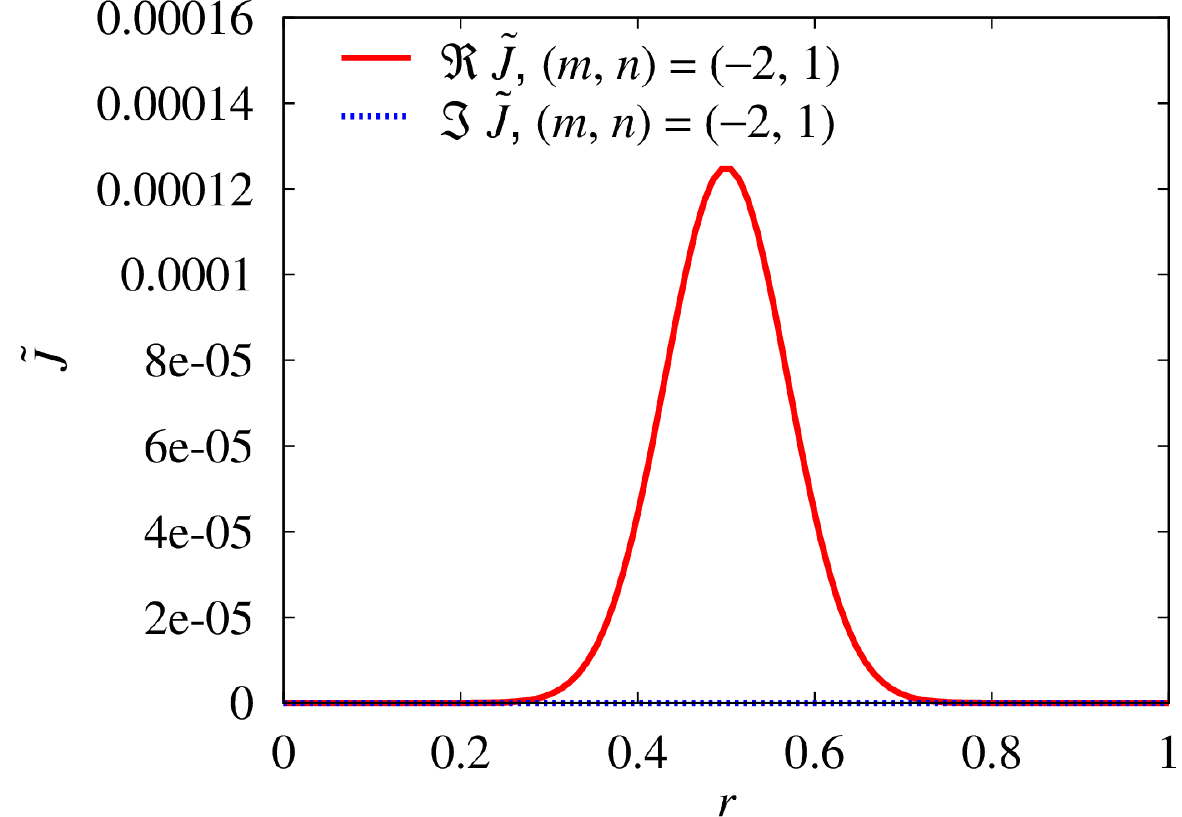}
 \subcaption{$\tilde{J}$. Imaginary part is zero.}
 \label{fig:r-tJ-perturb}
\end{minipage} 
 \caption{Radial profiles of advection fields $\tilde{\vphi}$ and $\tilde{J}$ used for generating dynamically accessible perturbations, i.e.,  perturbations on the same Casimir leaf as the equilibrium under consideration.}
 \label{fig:r-tvphi-tJ-perturb}
\end{figure*}

The equilibrium in the previous subsection~\ref{subsubsec:equilibrium} is perturbed by using arbitrary advection fields, as explained in Sec.~\ref{sec:procedure}.  We  choose the advection fields to be
\begin{align}
 \tilde{\vphi}(r, \theta, \zeta)
&=
A_{\vphi} r (1-r) 
\rme^{-\left( \frac{r-r_{0}}{L} \right)^{2}}
 \sin(2 \theta - \zeta),
 \label{eq:tvphi-perturb}
 \\
 \tilde{J}(r, \theta, \zeta)
&=
A_{J} r (1-r) 
\rme^{-\left( \frac{r-r_{0}}{L} \right)^{2}}
\cos (2\theta - \zeta),
 \label{eq:tJ-perturb}
\end{align}
where $A_{\vphi} = A_{J} = 10^{-3}$, $r_{0} = 0.5$ and $L = 0.1$ are all
fixed in time.   The Fourier-decomposed radial profiles with this procedure are  shown in Fig.~\ref{fig:r-tvphi-tJ-perturb}.

The initial condition for the time evolution is the cylindrically symmetric equilibrium. Thus the $(m, n) = (-2, 1)$ components of $U$ and $\psi$ are generated in the beginning phase just after the time evolution is started. Then the nonlinear terms between the generated $(m, n) = (-2, 1)$ components and the $(m, n) = (-2, 1)$ modes of the advection fields generate the $(m, n) = (-4, 2)$ and $(0, 0)$ components. When the   cylindrically symmetric equilibrium is perturbed in this way,  the result stays on the same Casimir leaf.

\begin{figure*}
\begin{minipage}[t]{0.45\textwidth}
 \centering
 \includegraphics[width=\textwidth]{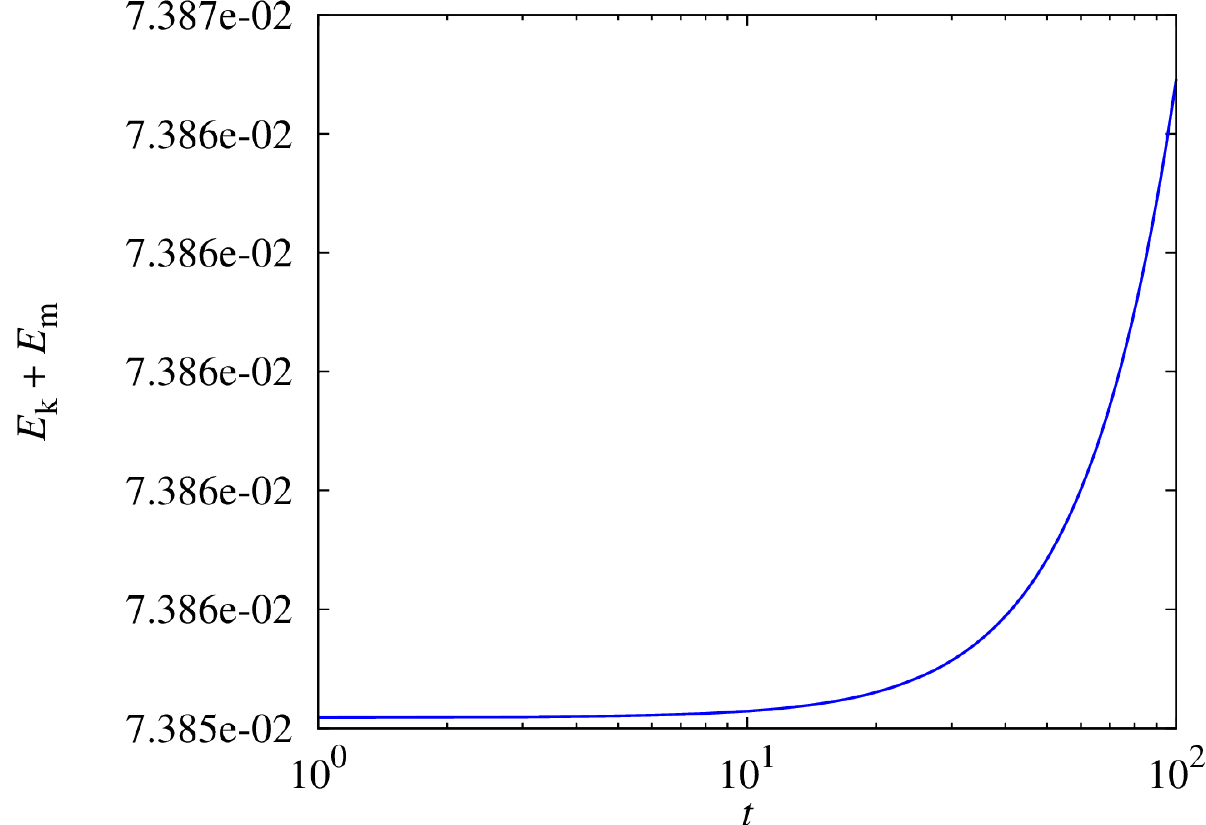}
 \subcaption{The total energy increases in time,
 where $E_{\rm k}$ and $E_{\rm m}$
 are kinetic and magnetic energies, respectively.}
 \label{fig:t-E-perturb}
\end{minipage} 
 \hfill
\begin{minipage}[t]{0.45\textwidth}
 \centering
 \includegraphics[width=\textwidth]{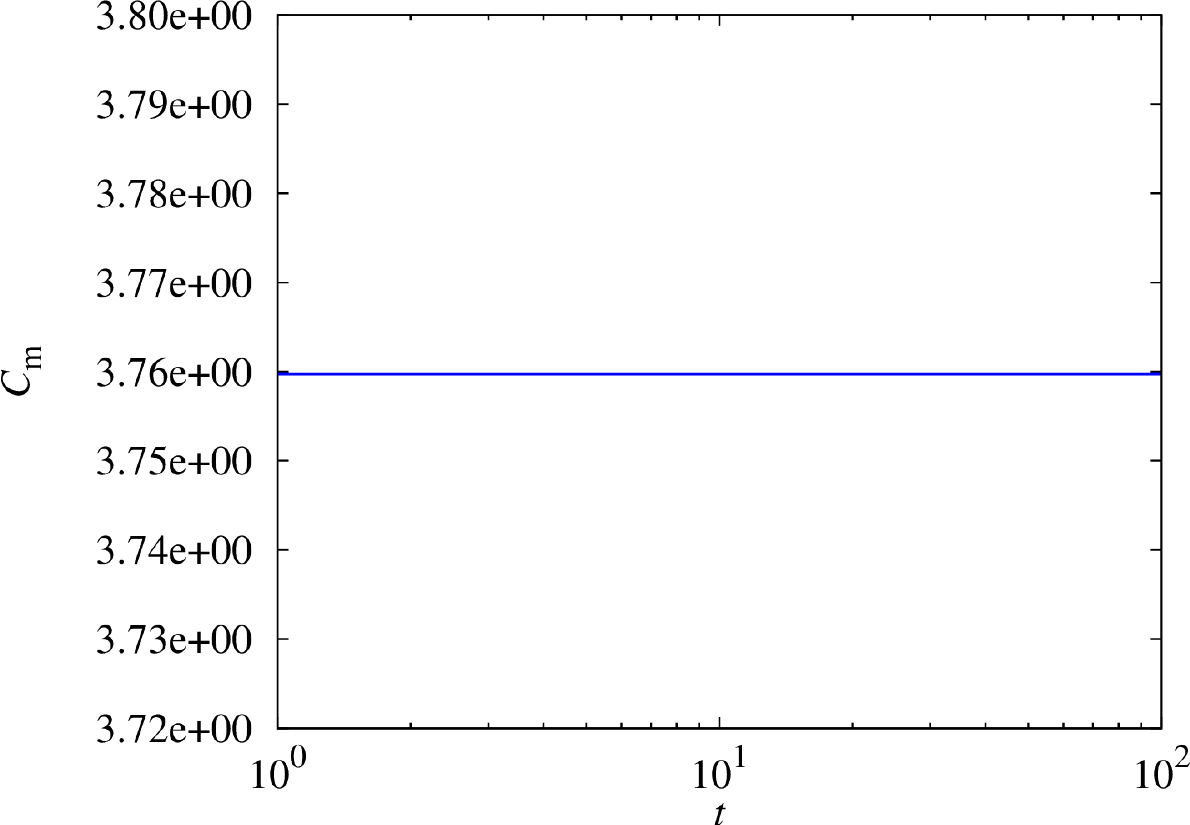}
 \subcaption{Magnetic helicity $C_{m}$ does not
 change in time.}
 \label{fig:t-Cm-perturb}
\end{minipage} 
 \caption{Time evolution of the total energy and the magnetic helicity, where the arbitrary chosen advection fields
 (\ref{eq:tvphi-perturb}) and (\ref{eq:tJ-perturb}) are used for perturbing the cylindrically symmetric equilibrium. }
 \label{fig:t-E-Cm-perturb}
\end{figure*}

The time evolution of the total energy,  the sum of the kinetic energy $E_{\rm k}$ and the magnetic energy $E_{\rm m}$, and the magnetic helicity $C_{\rm m}$, which is one of the Casimir invariants, are shown in Fig.~\ref{fig:t-E-Cm-perturb}.  When the arbitrarily chosen advection fields in Eqs.~(\ref{eq:tvphi-perturb}) and (\ref{eq:tJ-perturb}) are used for perturbing the cylindrically symmetric equilibrium, we observe that the total energy increases in time, while the magnetic helicity remains constant.

\begin{figure*}
\begin{minipage}[t]{0.45\textwidth}
 \centering
 \includegraphics[width=\textwidth]{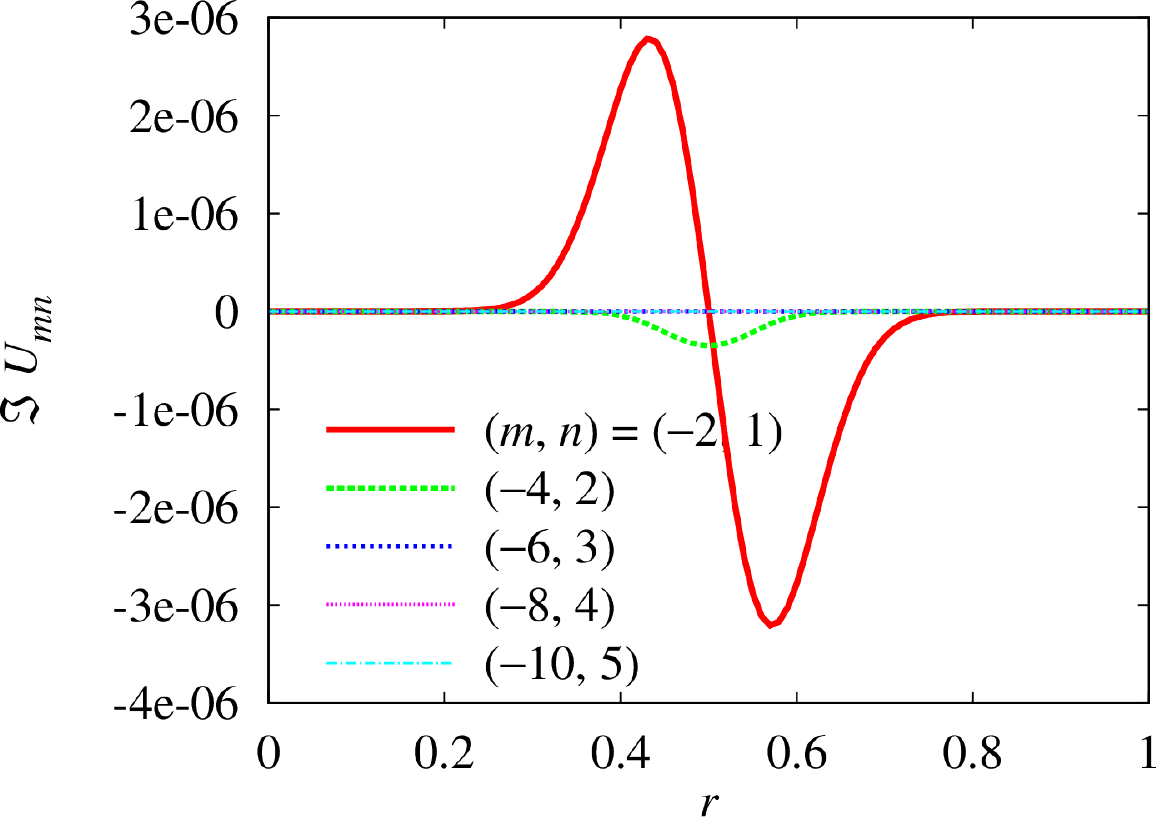}
 \subcaption{
 Radial profile of $\Im\, U_{mn}$.  The real part is zero.
 }
 \label{fig:r-U_i-t000000_0000}
\end{minipage} 
 \hfill
\begin{minipage}[t]{0.45\textwidth}
 \centering
 \includegraphics[width=\textwidth]{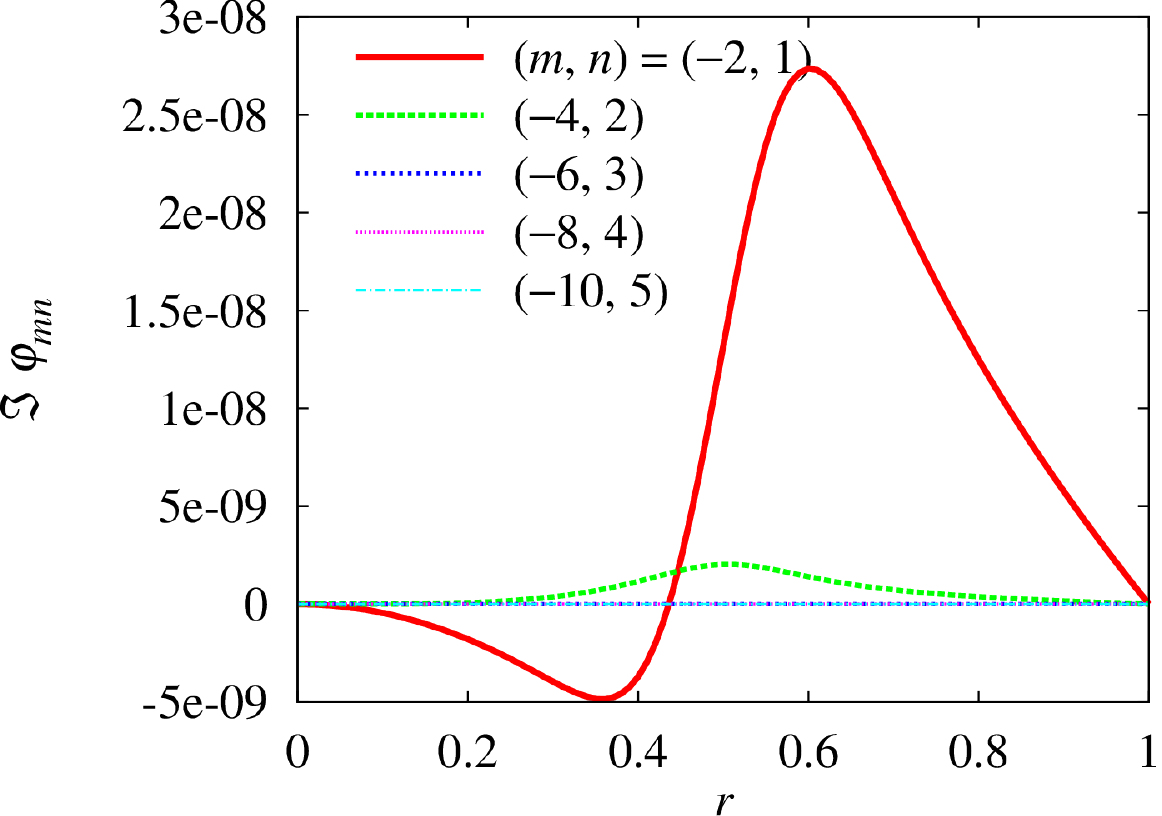}
 \subcaption{
 Radial profile of $\Im\, \vphi_{mn}$.  The real part is zero.
 }
 \label{fig:r-phi_i-t000000_0000}
\end{minipage} 
\begin{minipage}[t]{0.45\textwidth}
 \centering
 \includegraphics[width=\textwidth]{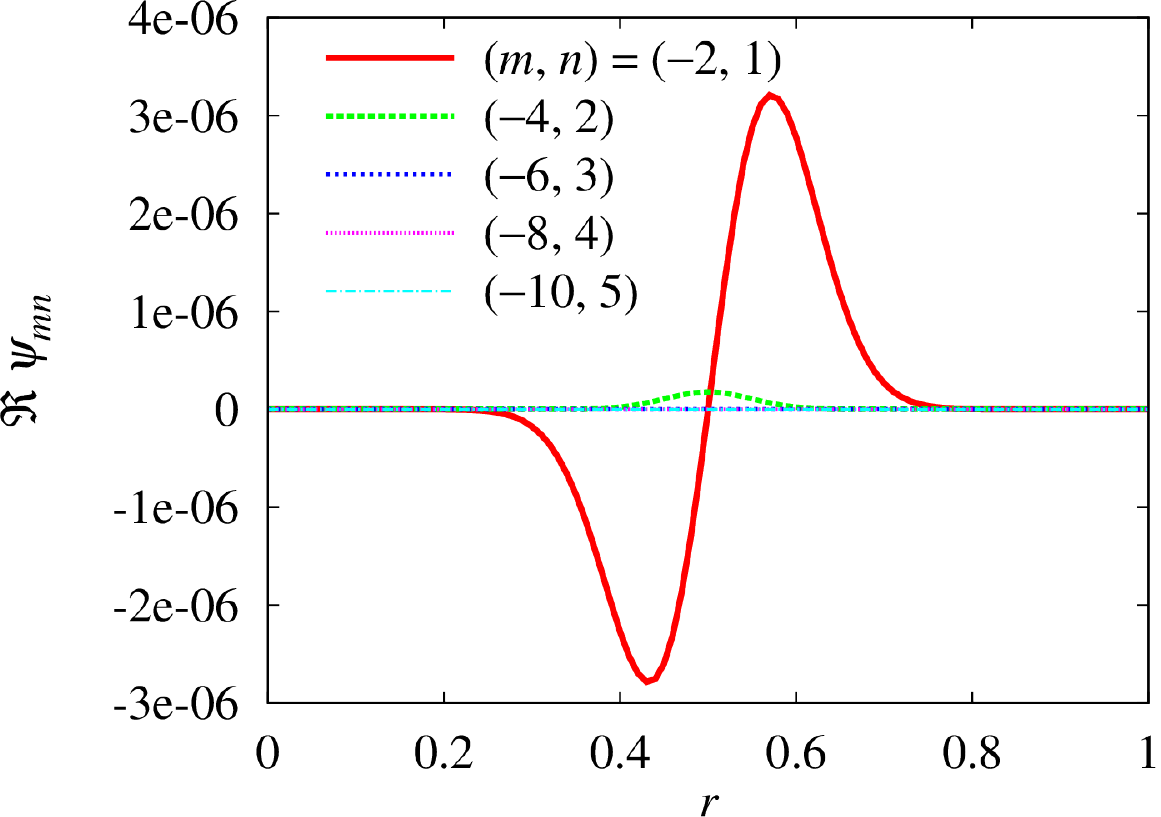}
 \subcaption{
 Radial profile of $\Re\, \psi_{mn}$.  The imaginary part is zero.
 }
 \label{fig:r-psi_r-t000000_0000}
\end{minipage} 
 \hfill
\begin{minipage}[t]{0.45\textwidth}
 \centering
 \includegraphics[width=\textwidth]{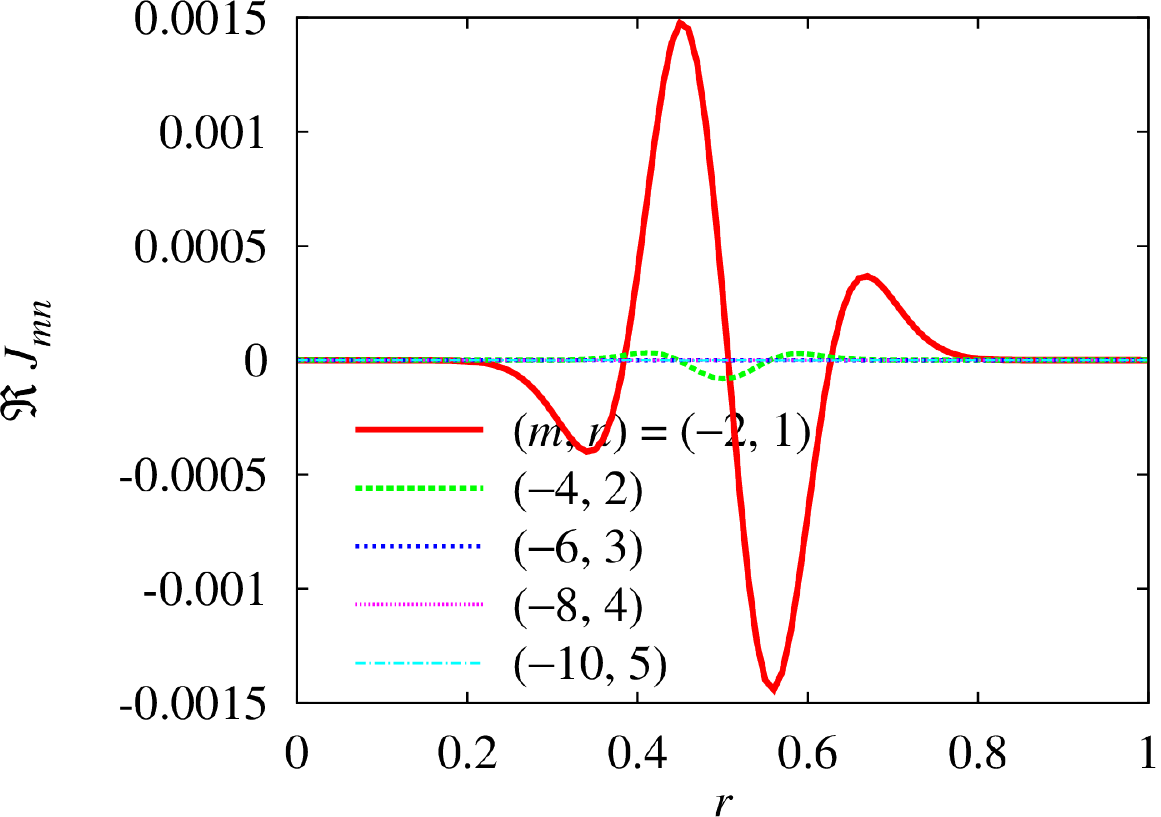}
 \subcaption{
 Radial profile of $\Re\, J_{mn}$.  The imaginary part is zero.
 }
 \label{fig:r-J_r-t000000_0000}
\end{minipage} 
 \caption{
 Radial profiles of the perturbed state that is used as an initial condition for  SA.
 The $(m, n) = (-2, 1)$ components are dominant, but they still have small amplitudes.
 }
 \label{fig:r-u-t000000_0000}
\end{figure*}

In the reminder of Sec.~\ref{sec:results}, we perform  SA to minimize
the total energy of the system by using the advection fields
Eqs.~(\ref{eq:tvphi}) and (\ref{eq:tJ}). The initial condition is chosen
as the state at $t=10$ in Fig.~\ref{fig:t-E-Cm-perturb}.
The radial profiles are shown in Fig.~\ref{fig:r-u-t000000_0000}.

\subsubsection{SA slow convergence of the velocity}
\label{subsubsec:slowConvergence}

\begin{figure*}
\begin{minipage}[t]{0.45\textwidth}
 \centering
 \includegraphics[width=\textwidth]{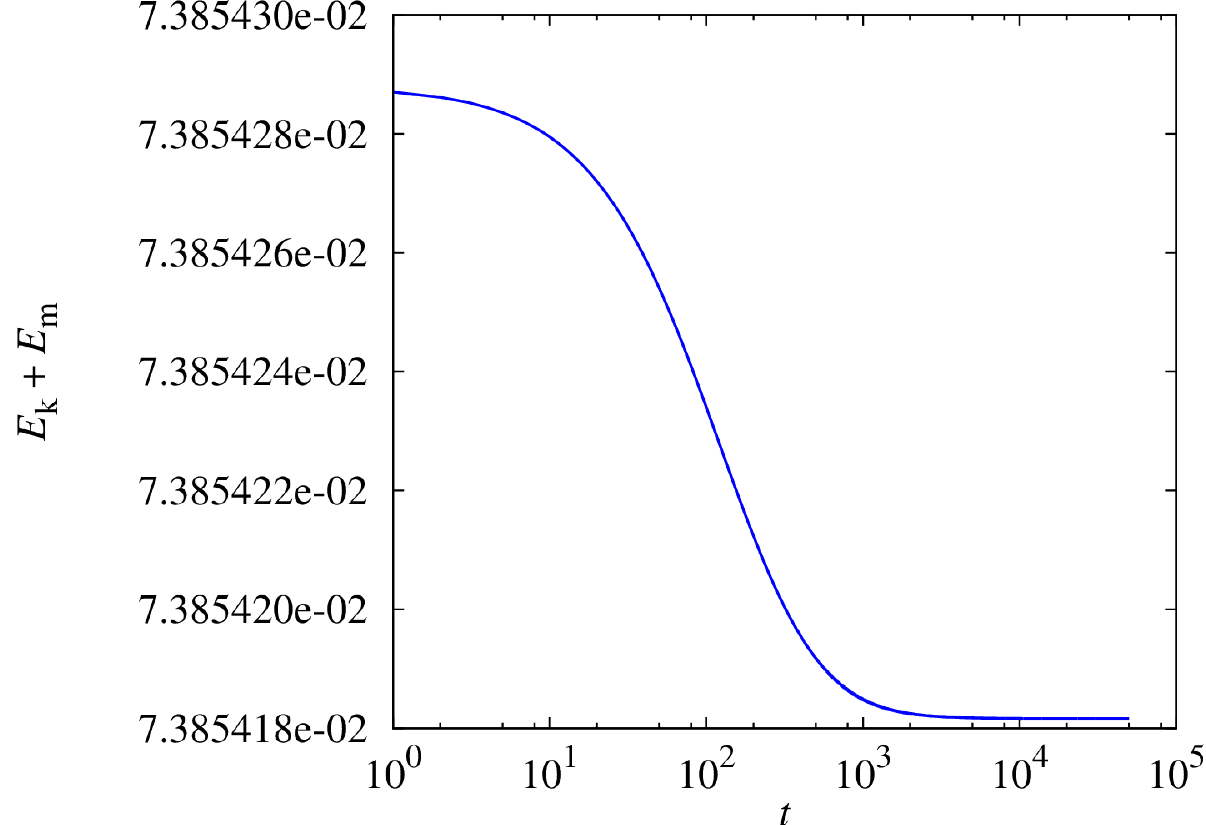}
 \subcaption{The total energy monotonically decreases in time.}
 \label{fig:t-E}
\end{minipage} 
 \hfill
\begin{minipage}[t]{0.45\textwidth}
 \centering
 \includegraphics[width=\textwidth]{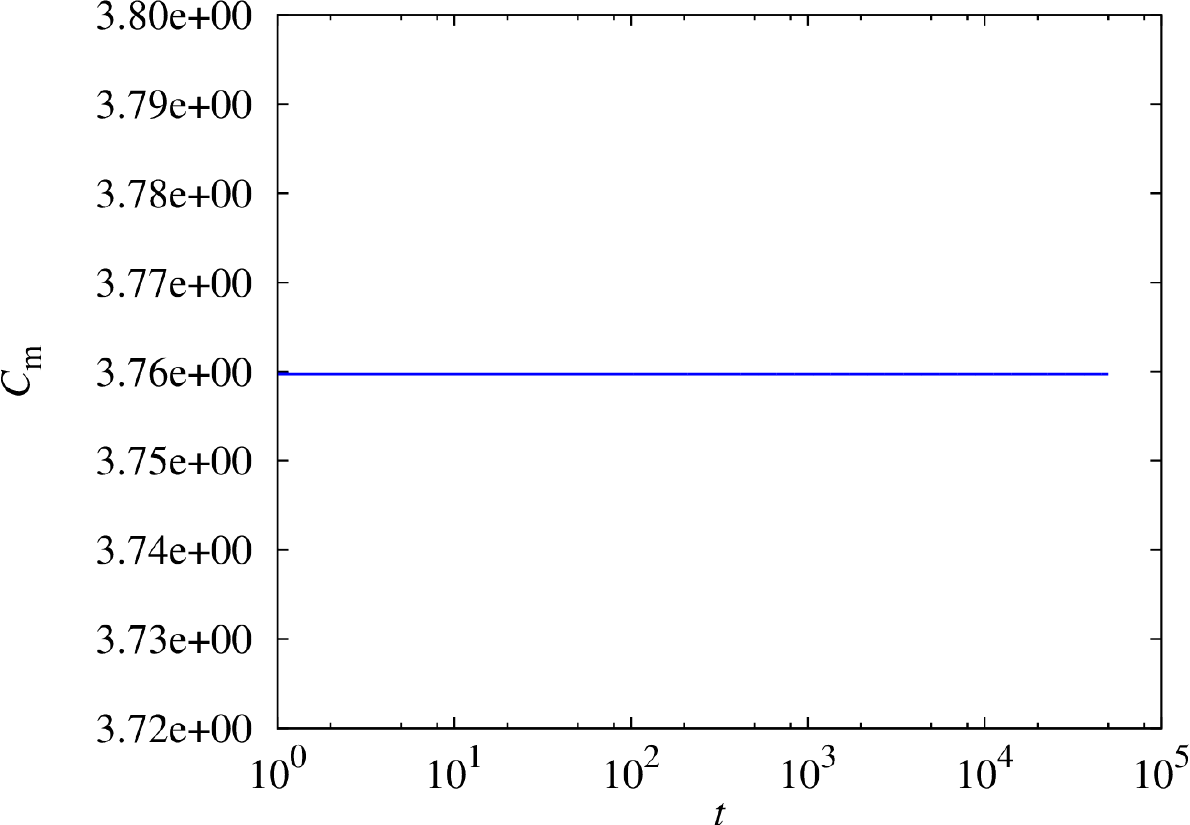}
 \subcaption{Magnetic helicity $C_{\rm m}$ does not
 change in time.}
 \label{fig:t-Cm}
\end{minipage} 
 \caption{Time evolution plots  of the total energy and  magnetic helicity
 during the SA evolution.}
 \label{fig:t-E-Cm}
\end{figure*}

Figure~\ref{fig:t-E-Cm} shows the time evolution of total energy and the magnetic helicity during the SA evolution. Observe, the total energy monotonically decreases until the system appears to reach a steady state when the energy remains nearly constant for a long time.  Note that the horizontal axis is a log scale.  Therefore, the state seems to be an equilibrium that has the minimum energy.  Also, Fig.~\ref{fig:t-E-Cm} shows that the magnetic helicity does not change during the SA evolution.

\begin{figure*}
\begin{minipage}[t]{0.45\textwidth}
 \centering
 \includegraphics[width=\textwidth]{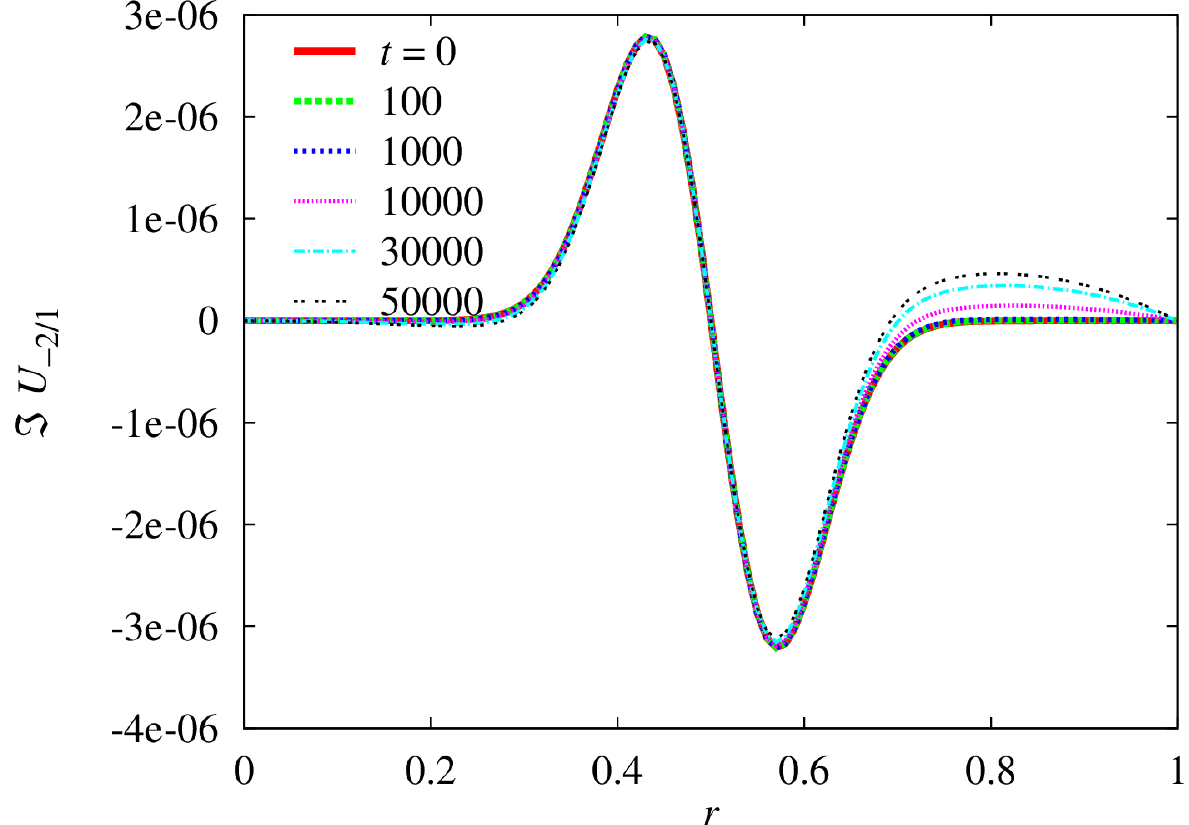}
 \subcaption{
 Radial profile of $\Im\, U_{-2,1}$.
 }
 \label{fig:r-U_i-t-mm2n1}
\end{minipage} 
 \hfill
\begin{minipage}[t]{0.45\textwidth}
 \centering
 \includegraphics[width=\textwidth]{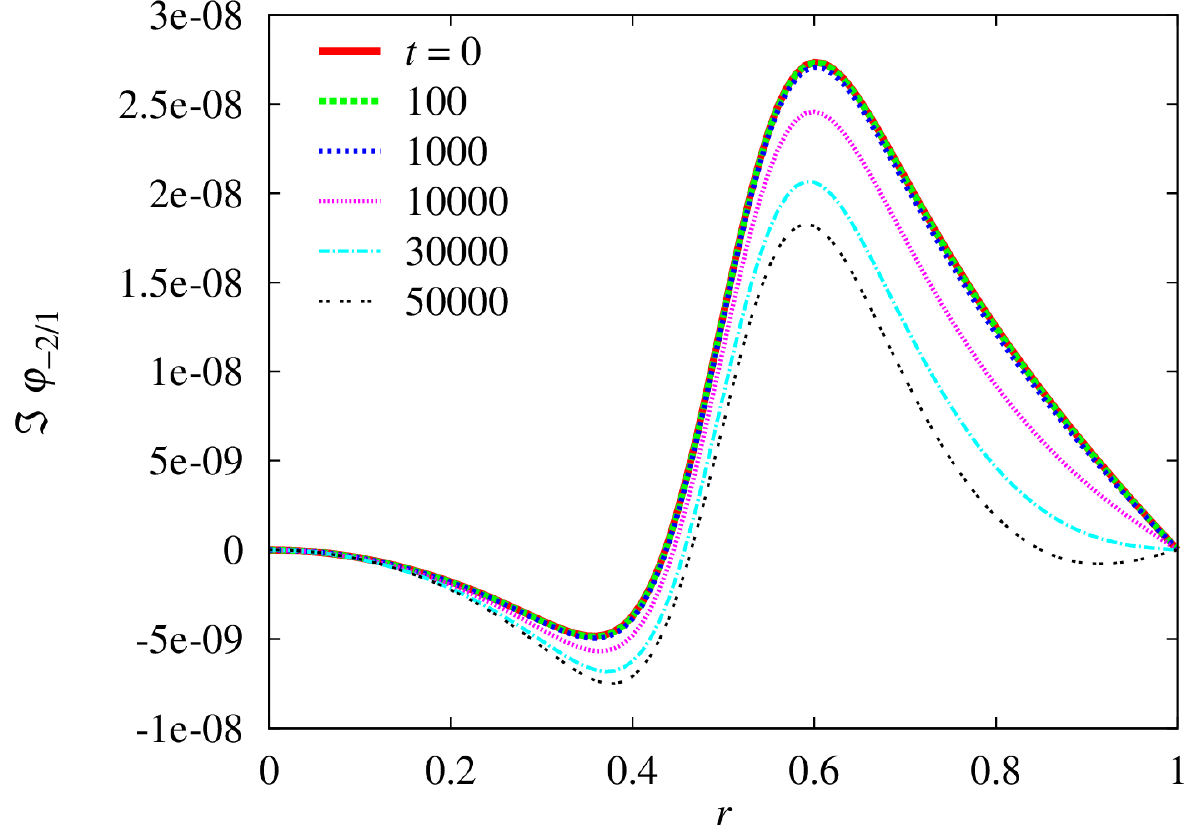}
 \subcaption{
 Radial profile of $\Im\, \vphi_{-2,1}$.
 }
 \label{fig:r-phi_i-t-mm2n1}
\end{minipage} 
\begin{minipage}[t]{0.45\textwidth}
 \centering
 \includegraphics[width=\textwidth]{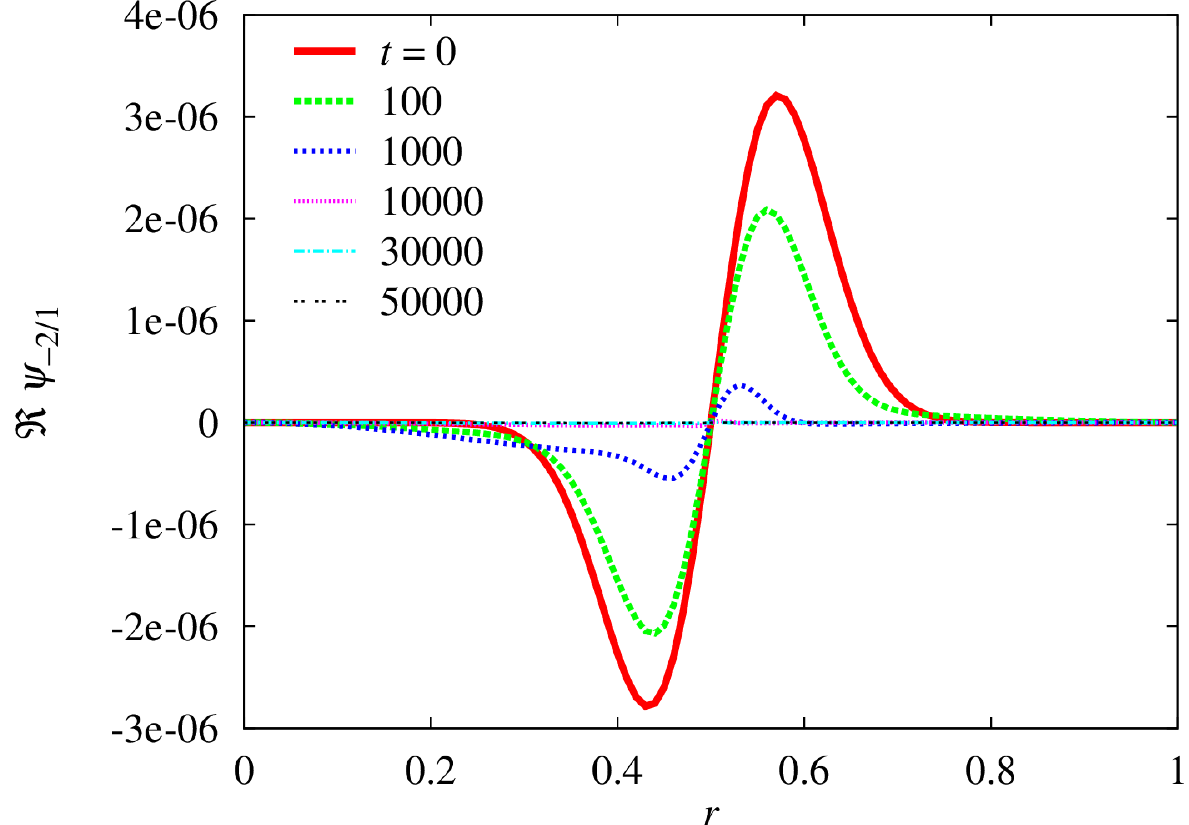}
\subcaption{
 Radial profile of $\Re\, \psi_{-2,1}$.
 zero.
 }
 \label{fig:r-psi_r-t-mm2n1}
\end{minipage} 
 \hfill
\begin{minipage}[t]{0.45\textwidth}
 \centering
 \includegraphics[width=\textwidth]{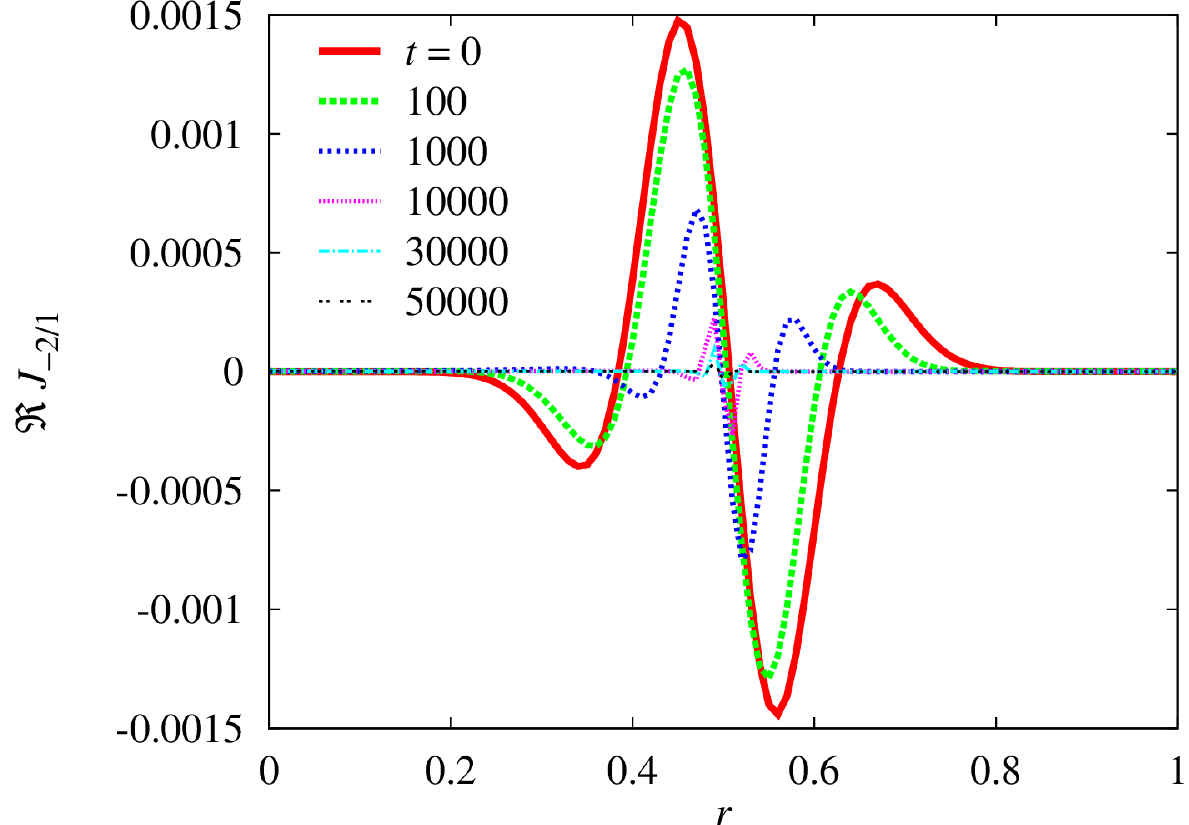}
 \subcaption{
 Radial profile of $\Re\, J_{-2,1}$.
 }
 \label{fig:r-J_r-t-mm2n1}
\end{minipage} 
 \caption{
 Radial profiles of the $(m, n) = (-2, 1)$ components are plotted  at several times during the SA evolution. The real parts of $U_{-2,1}$ and $\vphi_{-2,1}$ and the imaginary parts of $\psi_{-2,1}$ and $J_{-2,1}$ remain zero.  
 The magnetic part $\Re \,\psi_{-2,1}$ and $\Re\,J_{-2,1}$ disappear relatively quickly, while the kinetic part $\Im \,U_{-2,1}$ and
 $\Im \,\vphi_{-2,1}$ are changing in the almost stationary state of the energy.
 }
 \label{fig:r-u-t-mm2n1}
\end{figure*}

Figure~\ref{fig:r-u-t-mm2n1} shows the  time evolution of radial profiles of the $(m, n) = (-2, 1)$ components of \subref{fig:r-U_i-t-mm2n1} $\Im \,U$, \subref{fig:r-phi_i-t-mm2n1} $\Im \,\vphi$, \subref{fig:r-psi_r-t-mm2n1} $\Re \,\psi$ and \subref{fig:r-J_r-t-mm2n1} $\Re \,J$, respectively. Note that $\Re \,U_{-2,1}$, $\Re \,\vphi_{-2,1}$,  $\Im \,\psi_{-2,1}$ and $\Im \,J_{-2,1}$ remain almost zero during  SA.   Even though the total energy starts to decrease just after the SA is started, the velocity components do not change visibly in the early stage. The change of the velocity components become visible even after  the total energy reaches the stationary state, with these components changing even at $t=5 \times 10^{4}$. On the other hand, the magnetic components start to decrease in the early stage of  SA. Although $\Re \,J_{-2,1}$ still remains finite at $t=3 \times 10^{4}$, the magnetic components are negligibly small at $t = 5 \times 10^{4}$.  Thus, the rates of change  of the velocity and magnetic components are significantly different.

This may be explained as follows.  It is firstly pointed out that the $(m, n) = (-2, 1)$ components of $U$ and $\psi$,
the dominant Fourier modes of the perturbation, are ${\cal O}(10^{-6})$.  Then $\vphi$ is ${\cal O}(10^{-8})$, while $J$ is ${\cal O}(10^{-3})$.  This is because $\vphi$ is obtained by integrating $U$ especially in the radial direction, while $J$ is obtained by differentiating $\psi$. Let us consider the very early stage of  SA.  The $(m, n) = (-2, 1)$ component of $f^{1}$, or the right-hand sides of Eq.~(\ref{eq:vorticity-equation}), is generated mainly by the $[\psi_{00}, J_{-2,1}]$ term. On the other hand, the 
$(m, n) = (-2, 1)$ component of $f^{2}$, or the right-hand sides of Eq.~(\ref{eq:Ohm-law}), is generated by the $[\psi_{00}, \vphi_{-2,1}]$ term. Thus the $(m, n) = (-2, 1)$ component of $f^{1}$ is much larger than that of $f^{2}$.

The artificial advection fields $\tilde{\vphi}$ and $\tilde{J}$ are given by Eqs.~(\ref{eq:tvphi}) and (\ref{eq:tJ}), respectively.
Since $(K_{ij})$ is taken to be diagonal in this study, the $(m, n) = (-2, 1)$ component of  $\tilde{\vphi}$ is much larger than that of $\tilde{J}$. Then, in the SA equations (\ref{eq:SA-vorticity-equation}) and (\ref{eq:SA-Ohm-law}), the $[\psi_{00}, \tilde{\vphi}_{-2,1}]$ term in Eq.~(\ref{eq:SA-Ohm-law})  is much larger than the $[\psi_{00}, \tilde{J}_{-2,1}]$ term in  Eq.~(\ref{eq:SA-vorticity-equation}). Therefore $\psi_{-2,1}$ starts to change much faster than $U_{-2,1}$. Note that the $[U, \tilde{\vphi}]$ term is negligible especially in the early stage of the SA since $U_{00}$ is zero initially.

The SA equations for  reduced MHD have two advection fields. Therefore, depending on the relative magnitudes, the relaxation path can change.  In the present settings, the relaxation of the magnetic component is much faster than the velocity component.
This may be understood schematically by supposing  the energy level corresponds to the height of a mountain with topography.
Then, the perturbed state corresponds to a point somewhere on a hillside of the mountain and the equilibrium is the place of the lowest height. There can be a variety of paths to the equilibrium.  In the present case, the system goes down to the valley quickly, and then moves slowly along a valley to the place of the lowest height. Thus, the question is whether the path can be controlled without losing the characteristics of  SA:  monotonic change of the energy while preserving the Casimir invariants. This would be realized by forcing the two advection fields to have same order of magnitudes, which is what we will  propose in Sec.~\ref{subsubsec:acceleration}.

\subsubsection{SA relaxation with a forbidden direction}
\label{subsubsec:forbiddenDirection}

Before devising the advection fields that will speed up SA relaxation, let us first examine what happens if one of the advection fields is fully dropped. Since the system loses one of its fields for the relaxation, we expect  the relaxation to be  incomplete.
Dropping one of the fields can be accomplished by setting either $\alpha_{11} = 0$ or $\alpha_{22} = 0$.

\begin{figure*}
\begin{minipage}[t]{0.45\textwidth}
 \centering
 \includegraphics[width=\textwidth]{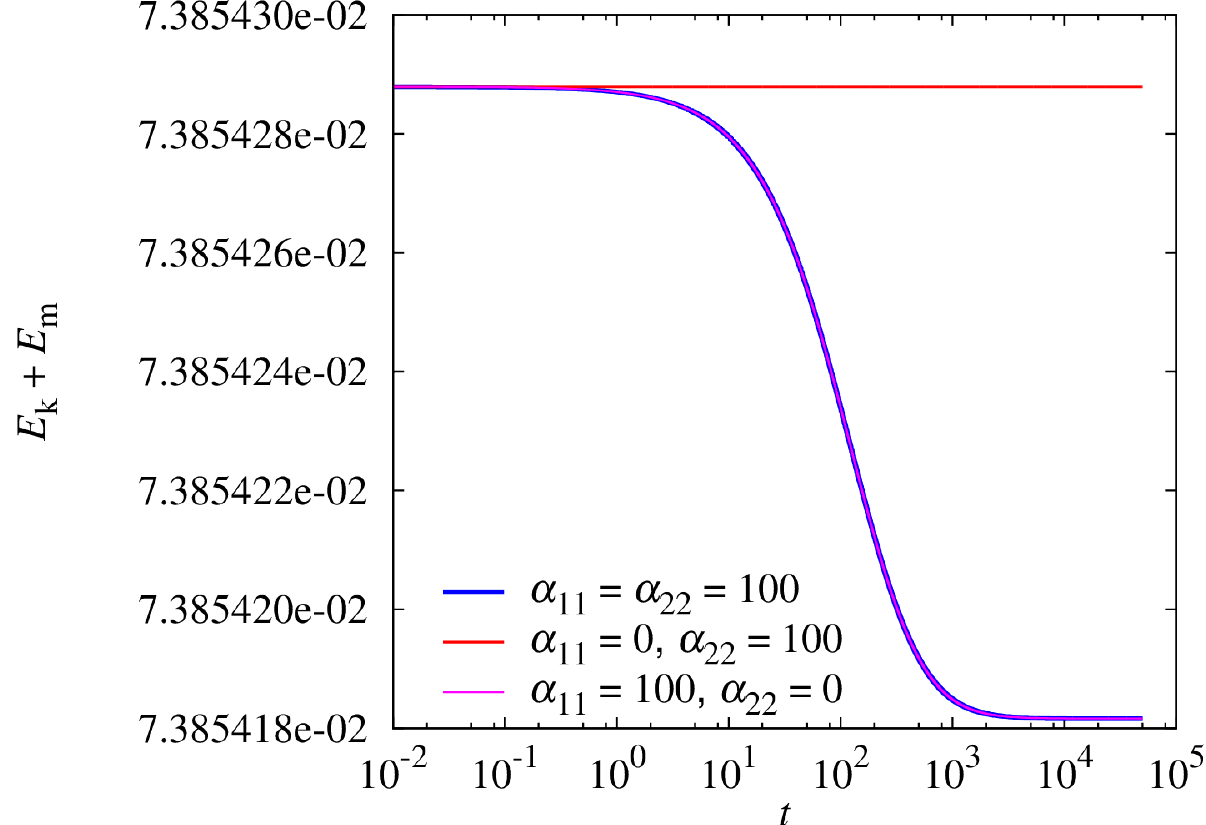}
 \subcaption{
 Total energy.
 }
 \label{fig:t-E-w_alpjj-0}
\end{minipage} 
 \\
\begin{minipage}[t]{0.45\textwidth}
 \centering
 \includegraphics[width=\textwidth]{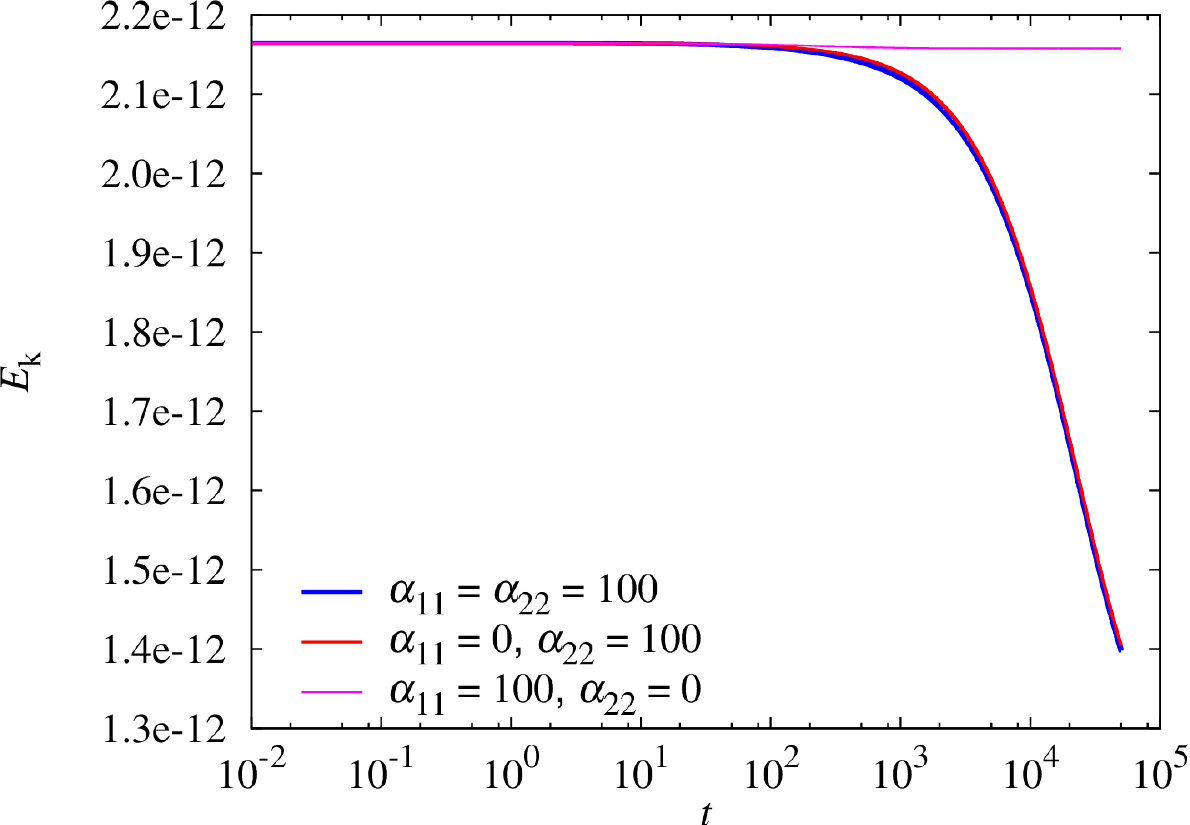}
 \subcaption{
 Kinetic energy.
 }
 \label{fig:t-Ek-w_alpjj-0}
\end{minipage} 
 \hfill
\begin{minipage}[t]{0.45\textwidth}
 \centering
 \includegraphics[width=\textwidth]{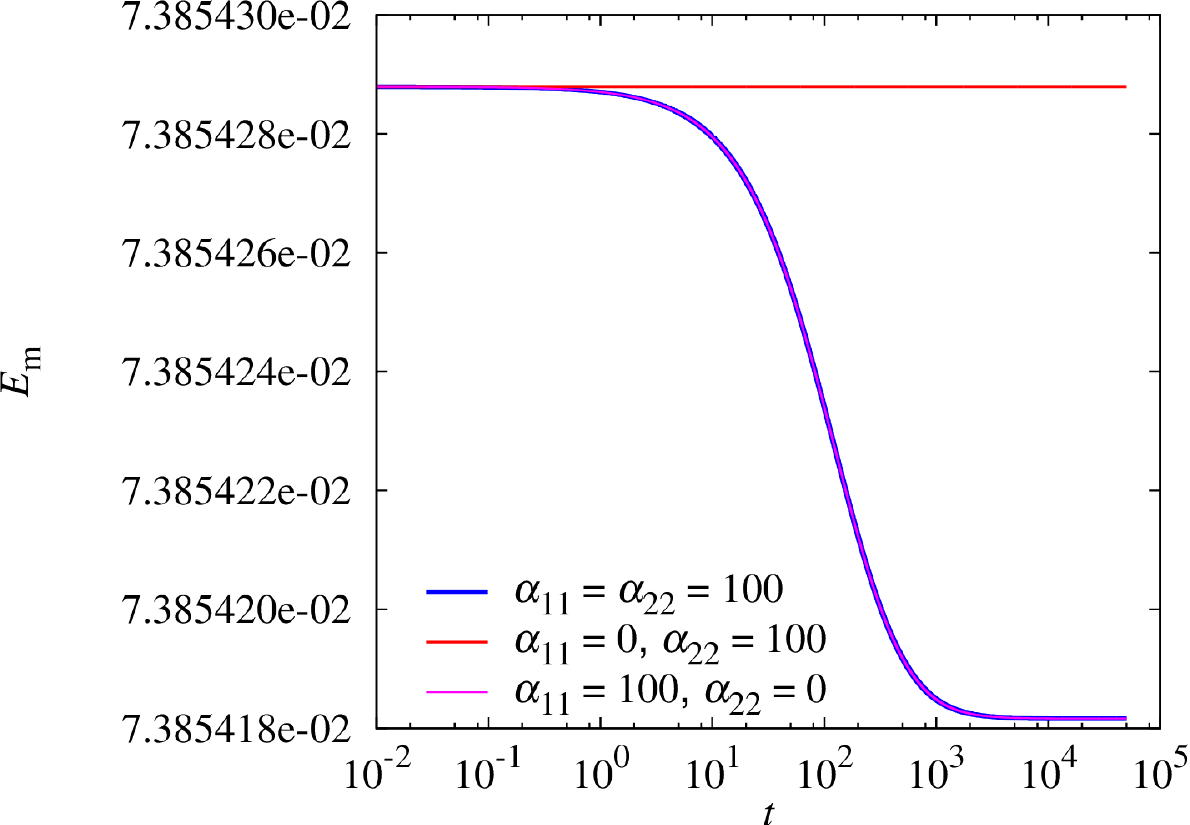}
 \subcaption{
 Magnetic energy.
 }
 \label{fig:t-Em-w_alpjj-0}
\end{minipage} 
 \caption{
 Time evolution of \subref{fig:t-E-w_alpjj-0} total energy,
 \subref{fig:t-Ek-w_alpjj-0} kinetic energy and
 \subref{fig:t-Em-w_alpjj-0} magnetic energy.
 If $\tilde{\vphi}$ is dropped by setting
 $\alpha_{11} = 0$, the dominant magnetic energy does not change and
 thus no relaxation occurs.  
 }
 \label{fig:t-E-Ek-Em-w_alpjj-0}
\end{figure*}

Figure~\ref{fig:t-E-Ek-Em-w_alpjj-0} shows the time evolution of \subref{fig:t-E-w_alpjj-0} total energy, \subref{fig:t-Ek-w_alpjj-0} kinetic energy and  \subref{fig:t-Em-w_alpjj-0} magnetic energy, respectively. The $\alpha_{11} = \alpha_{22} = 100$, the case of  Fig.~\ref{fig:t-E-Cm},  is plotted here for comparison.   Note that Fig.~\ref{fig:t-E-Ek-Em-w_alpjj-0} is plotted onward from $t=10^{-2}$ for  showing clearly that initial energy is the same for all cases.

In the case of $\alpha_{11} = 0$ and $\alpha_{22} = 100$, $\tilde{\vphi}$ becomes zero.  Then the magnetic energy does not change in time as shown in Fig.~\ref{fig:t-E-Ek-Em-w_alpjj-0}\subref{fig:t-Em-w_alpjj-0}. This is clear from Eq.~(\ref{eq:SA-Ohm-law}); $\psi$ can only evolve with finite $\tilde{\vphi}$.  Although the kinetic energy changes slightly due to the finite $\tilde{J}$ as seen in Fig.~\ref{fig:t-E-Ek-Em-w_alpjj-0}\subref{fig:t-Ek-w_alpjj-0},  the total energy remains almost unchanged since the magnetic energy is dominant.

On the other hand,  in the case of $\alpha_{11} = 100$ and $\alpha_{22} = 0$, $\tilde{J}$ becomes zero.  Then, the  magnetic energy changes in time due to finite $\tilde{\vphi}$.  The kinetic energy remains almost unchanged as seen in Fig.~\ref{fig:t-E-Ek-Em-w_alpjj-0}\subref{fig:t-Ek-w_alpjj-0}, although it can change slowly due to the nonlinear term  $[U, \tilde{\vphi}]$  in Eq.~(\ref{eq:SA-vorticity-equation}).  Since the relaxation of the dominant magnetic energy occurs in this case, the time evolution of the total energy almost overlaps the case of $\alpha_{11} = \alpha_{22} = 100$.

Summarizing, the magnetic energy can never decrease if $\tilde{\vphi}$ is completely dropped;  thus the system does not relax to a
minimum energy state.  In the case of vanishing $\tilde{J}$, the system can relax, however, it must be quite slow. Therefore, both advection fields should remain finite and of  comparable magnitude for obtaining efficient relaxation. This is the task addressed in Sec.~\ref{subsubsec:acceleration}.

\subsubsection{SA accelerated  relaxation}
\label{subsubsec:acceleration}

Let us now consider a method for maintaining  the advection fields to be of the same order of magnitude.
In this study, we do this by introducing time dependence in the quantity $(K_{ij})$ of  Eqs.~(\ref{eq:tvphi}) and (\ref{eq:tJ}),  while retaining positive definiteness.  Explicitly, we choose
\begin{align}
 \tilde{\vphi}(\vx, t)
 &:=
 \alpha_{11}(t)
 \int_{\cal D} \td^{3} x^{\pr} \,
 g(\vx, \vx^{\pr}) f^{1}(\vx^{\pr}, t),
\label{eq:tvphi-tdep}
 \\
 \tilde{J}(\vx, t)
 &:=
 \alpha_{22}(t)
 \int_{\cal D} \td^{3} x^{\pr} \,
 g(\vx, \vx^{\pr}) f^{2}(\vx^{\pr}, t)\,,
\label{eq:tJ-tdep}
\end{align}
where $\alpha_{jj}(t)$ are chosen according to 
\begin{equation}
 \alpha_{jj}(t)
=
\min
\left\{
\frac{F_{\rm max}}
     {\dis{\max_{r, m, n}}\, | f^{j}_{mn}(r, t) |}
\ 
,
\ 
\alpha_{\rm max}
\right\}. 
\label{eq:alp-tdep}
\end{equation}
Here,   $\max_{r, m, n} | f^{j}_{mn}(r, t) |$ is the maximum absolute value of the Fourier components of $f^{j}$ in $r$ and $F_{\rm max}$ is the desired magnitude of the advection field. The parameter $F_{\rm max}$ is taken to be the same for $j=1$ and $2$ for
balancing the order of magnitude of $\tilde{\vphi}$ and $\tilde{J}$.  Therefore, the first term in the curly braces controls the magnitude of the advection fields, forcing them  to have the essentially the same order of magnitude. The second term in the curly braces is introduced to avoid divergence of the $\alpha_{jj}$.  In the final phase of the relaxation, $\max_{r, m, n} | f^{j}_{mn}(r, t) |$ becomes vanishingly small and the upper limit on $\alpha_{\rm max}$  avoids the divergence caused by this.

\begin{figure*}
\begin{minipage}[t]{0.45\textwidth}
 \centering
 \includegraphics[width=\textwidth]{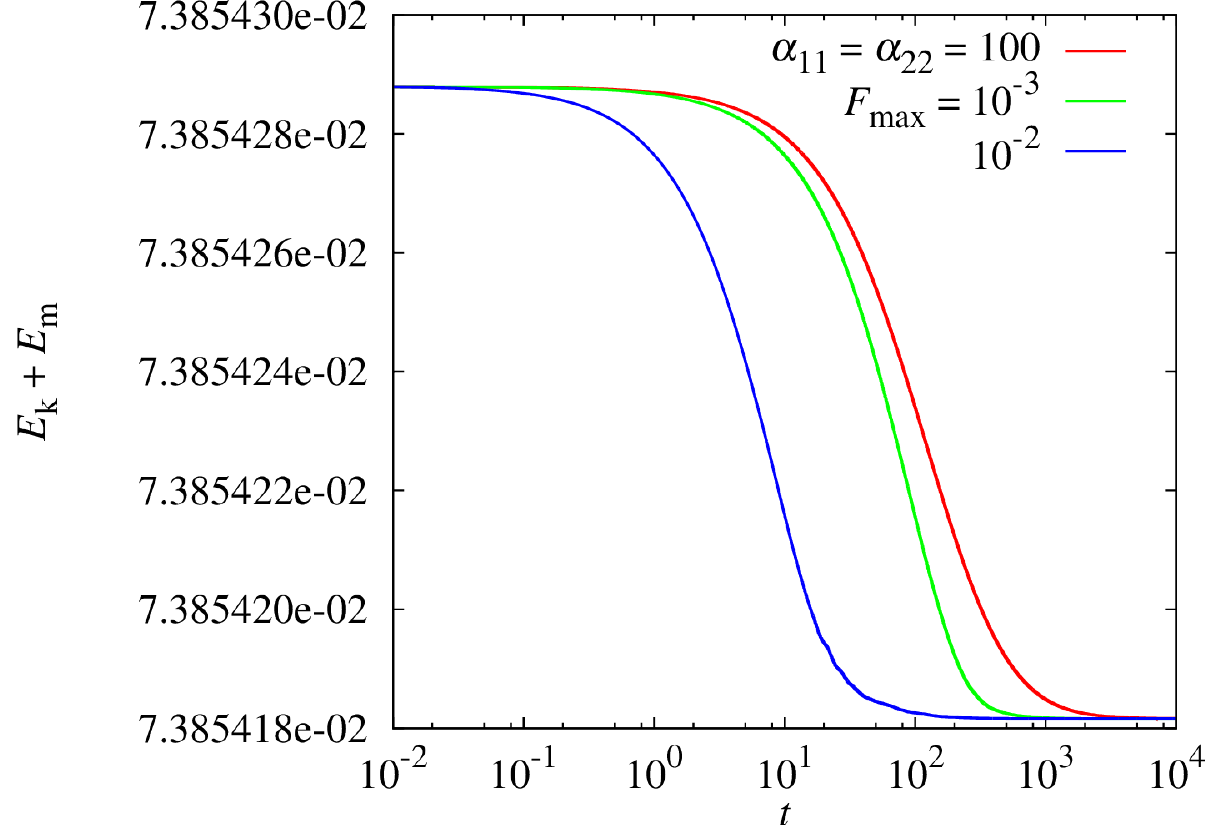}
 \subcaption{
 Total energy.
 }
 \label{fig:t-E-ralp}
\end{minipage}
 \hfill
\begin{minipage}[t]{0.45\textwidth}
 \centering
 \includegraphics[width=\textwidth]{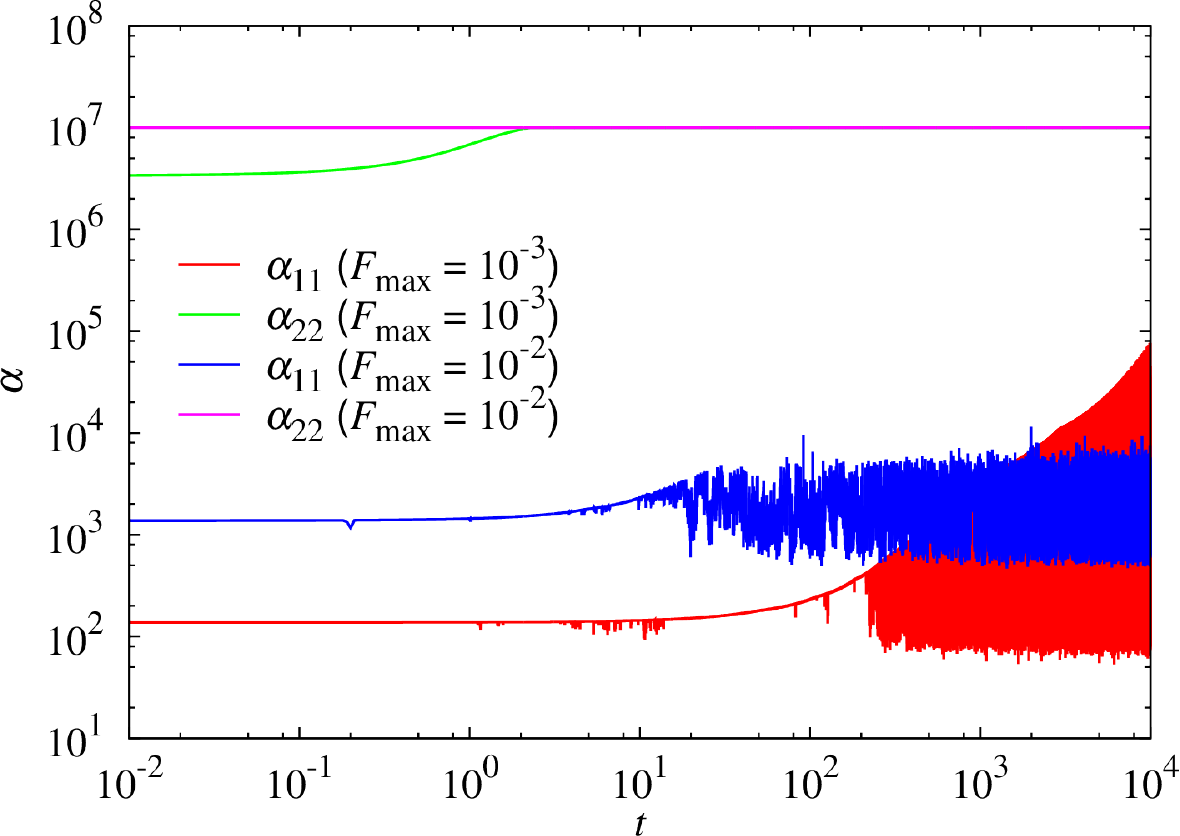}
 \subcaption{
 $\alpha_{11}$ and $\alpha_{22}$.
 }
 \label{fig:t-alp-ralp}
\end{minipage} 
\begin{minipage}[t]{0.45\textwidth}
 \centering
 \includegraphics[width=\textwidth]{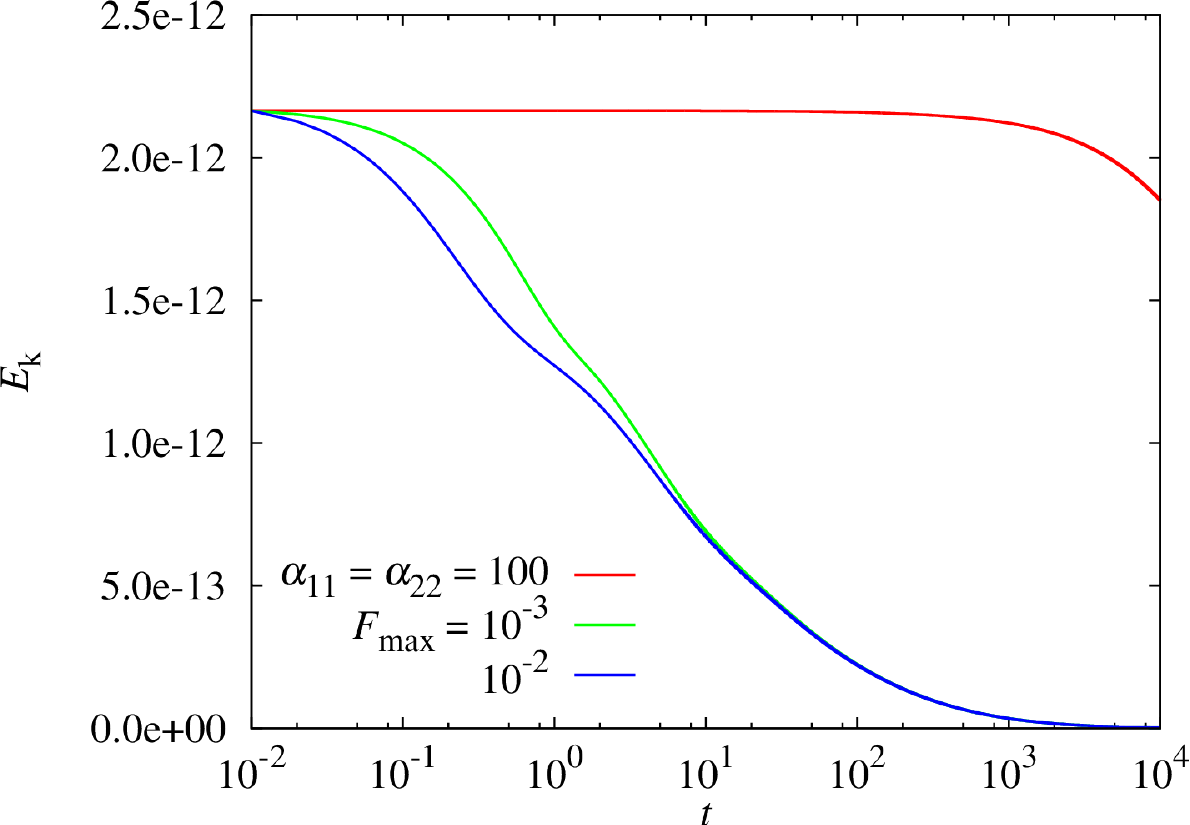}
 \subcaption{
 Kinetic energy.
 }
 \label{fig:t-Ek-ralp}
\end{minipage} 
 \hfill
\begin{minipage}[t]{0.45\textwidth}
 \centering
 \includegraphics[width=\textwidth]{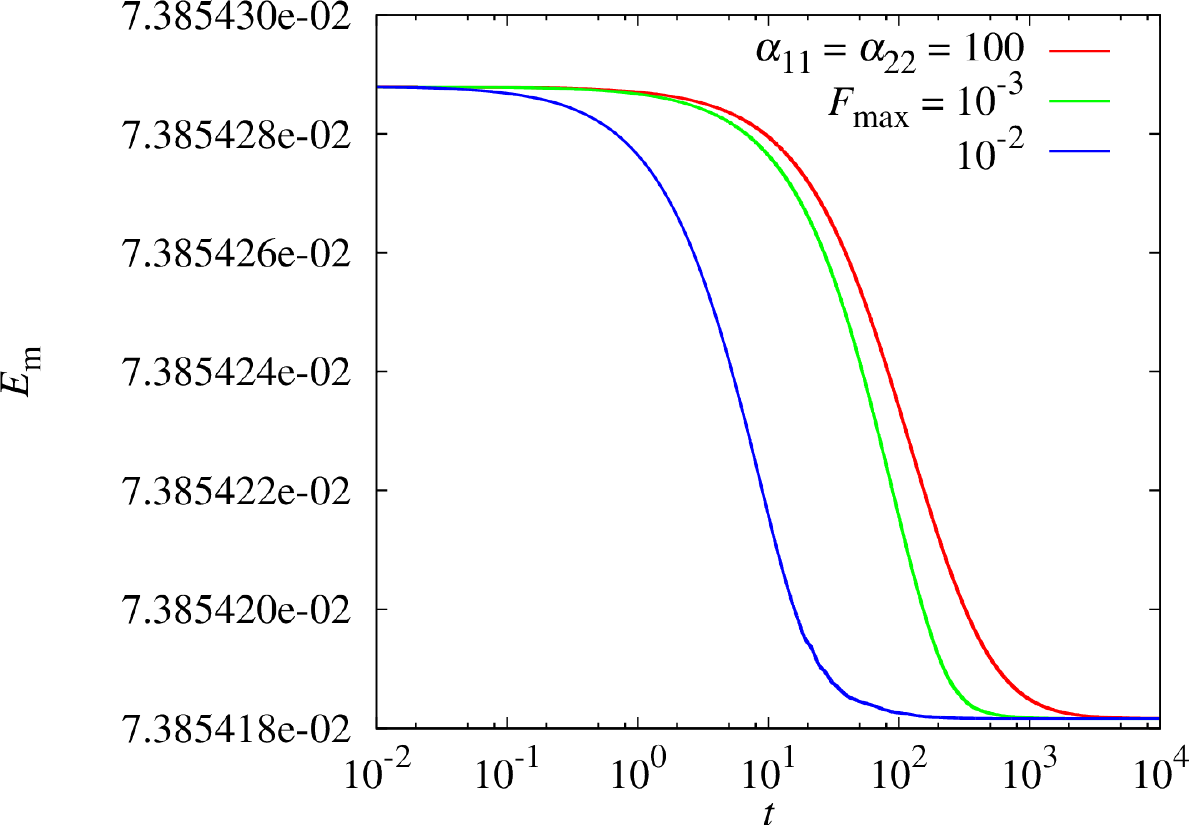}
 \subcaption{
 Magnetic energy.
 }
 \label{fig:t-Em-ralp}
\end{minipage} 
 \caption{
 Time evolution of
 \subref{fig:t-E-ralp} total energy,
 \subref{fig:t-alp-ralp} $\alpha_{11}$ and $\alpha_{22}$,
 \subref{fig:t-Ek-ralp} kinetic energy and 
 \subref{fig:t-Em-ralp} magnetic energy for
 $F_{\rm max} = 10^{-3}$ or $10^{-2}$.  The upper limit was set to
 $\alpha_{\rm max} = 10^{7}$.  The convergence to the stationary state
 is accelerated.
 }
 \label{fig:t-E-alp-Ek-Em-ralp}
\end{figure*}

Figure~\ref{fig:t-E-alp-Ek-Em-ralp} shows the time evolution of \subref{fig:t-E-ralp} total energy, \subref{fig:t-alp-ralp} $\alpha_{11}$ and $\alpha_{22}$, \subref{fig:t-Ek-ralp} kinetic energy and  \subref{fig:t-Em-ralp} magnetic energy for $F_{\rm max} = 10^{-3}$ and $10^{-2}$. The time evolution of energy without the time dependence of $\alpha_{jj}$, fixed $\alpha_{11} = \alpha_{22} = 100$, is also plotted in \subref{fig:t-E-ralp}, \subref{fig:t-Ek-ralp} and \subref{fig:t-Em-ralp}.   The upper limit for $\alpha_{jj}$ was set to
$\alpha_{\rm max} = 10^{7}$.   The convergence to the stationary state is clearly seen to be accelerated; for example,  when $F_{\rm max} = 10^{-2}$, the required time for reaching  the stationary state is considerably faster than the fixed $\alpha_{jj}$ case.

Especially, the time relaxation evolution of the kinetic energy
starts significantly earlier than the $\alpha_{11} = \alpha_{22} = 100$ case
as shown in Fig.~\ref{fig:t-E-alp-Ek-Em-ralp}\subref{fig:t-Ek-ralp}, although the order of magnitude of the kinetic energy is small.  In the case of fixed $\alpha_{jj}$, the kinetic energy starts to decrease even after reaching the almost stationary state of the total energy. On the other hand, the kinetic energy starts to decrease in the early stage of  SA when $\alpha_{jj}$ is varied in time.  

As seen in Fig.~\ref{fig:t-E-alp-Ek-Em-ralp}\subref{fig:t-alp-ralp},
$\alpha_{22}$ is significantly larger than $\alpha_{11}$.
This is because $f^{2}$ is much smaller than $f^{1}$.
Accordingly, the rapid decrease of the kinetic energy is obtained.

The choice of $F_{\rm max}$ and $\alpha_{\rm max}$ can largely affect the convergence efficiency. If $F_{\rm max}$ is too large, the simulation can diverge easily without an appropriate step-size control since
the right-hand sides of the evolution equations (\ref{eq:SA-vorticity-equation}) and (\ref{eq:SA-Ohm-law})
become too large.   A large $\alpha_{\rm max}$ also has similar tendency, although the effect on the divergence is indirect since $\alpha_{\rm max}$ is an upper limit for the right-hand sides of the evolution equation (\ref{eq:SA-vorticity-equation}) and (\ref{eq:SA-Ohm-law}).

Related to the convergence efficiency, there can be a better choice for $\alpha_{jj}$ than that of Eq.~(\ref{eq:alp-tdep}).  For example, one can expect defining $\alpha_{jj}$ for each $m$ and $n$ as $\alpha_{jj,mn}$.
However, this is not so straightforward since $\max_{r}\, |f_{mn}^{j}(r, t)|$ becomes smaller for larger $m$ and
$n$ and thus $\alpha_{jj,mn}$ can become huge or reach the upper limit. This means that the relative magnitude of the right-hand side associated with the evolved quantity $U_{mn}$ or $\psi_{mn}$ becomes large.  Then the simulation is likely to diverge. In this case, $F_{\rm max}$ and $\alpha_{\rm max}$ should be also chosen to have an appropriate order of magnitude for each $U_{mn}$ and $\psi_{mn}$. 

Another possible choice for accelerating convergence is to take the
symmetric kernel $(K_{ij})$ in Eqs.~(\ref{eq:tvphi}) and (\ref{eq:tJ})
with nonzero off-diagonal elements.  This mixes $f^{1}$ and $f^{2}$ and
can make the advection fields $\tilde{\vphi}$ and $\tilde{J}$ for SA to
be of comparable magnitudes.  We have not tried this choice yet.
Time dependence of the kernel may be necessary even with this choice to
accelerate the convergence near the energy minimum state.

\subsubsection{SA with another initial perturbation with larger 
   kinetic energy}
\label{subsubsec:anotherInitialPerturbation}

\begin{figure*}
\begin{minipage}[t]{0.45\textwidth}
 \centering
 \includegraphics[width=\textwidth]{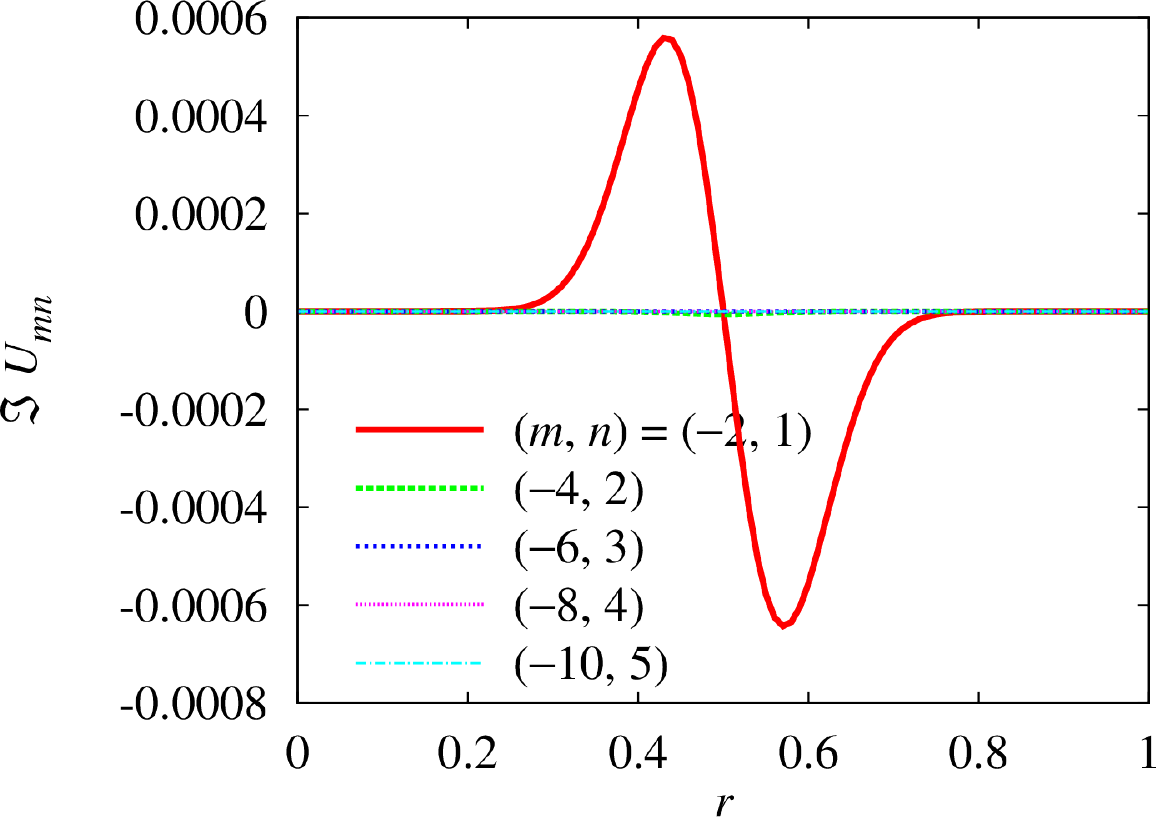}
 \subcaption{
 Radial profile of $\Im\, U_{mn}$.  The real part is zero.
 }
 \label{fig:r-U_i-t000000_0000-ini2}
\end{minipage} 
 \hfill
\begin{minipage}[t]{0.45\textwidth}
 \centering
 \includegraphics[width=\textwidth]{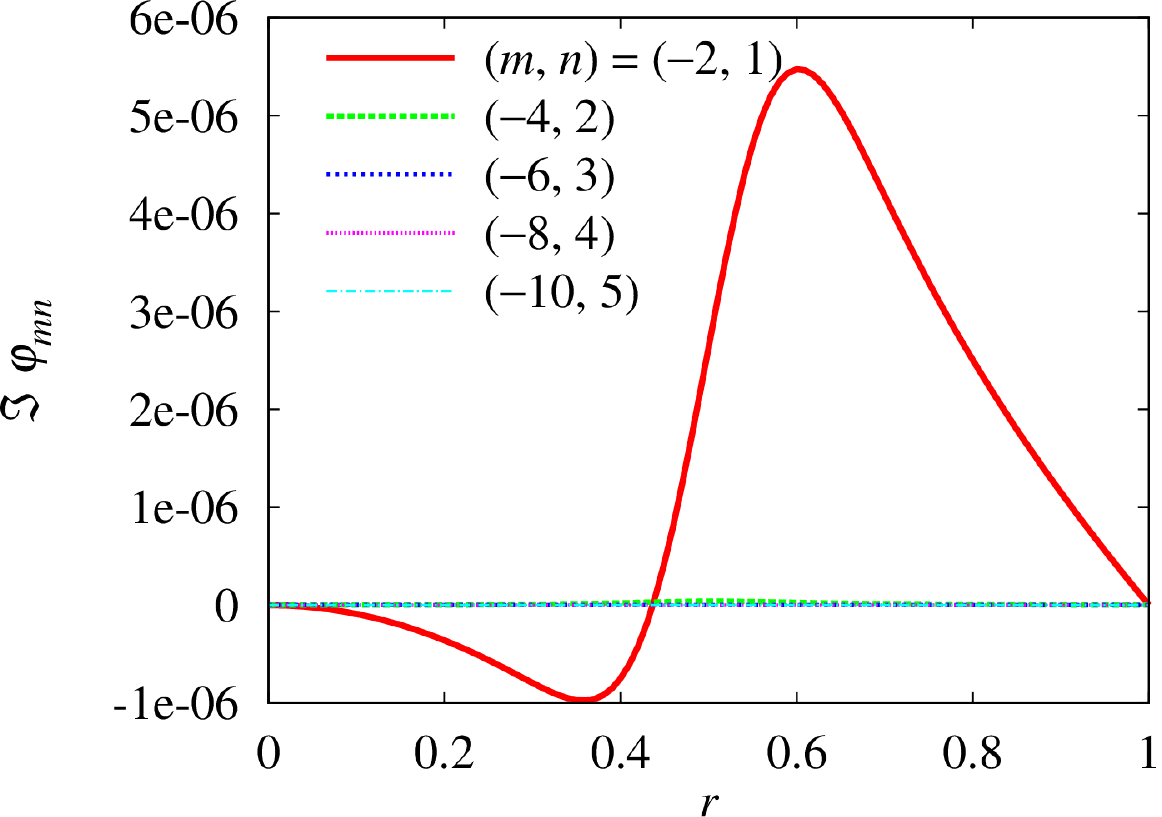}
 \subcaption{
 Radial profile of $\Im\, \vphi_{mn}$.  The real part is zero.
 }
 \label{fig:r-phi_i-t000000_0000-ini2}
\end{minipage} 
\begin{minipage}[t]{0.45\textwidth}
 \centering
 \includegraphics[width=\textwidth]{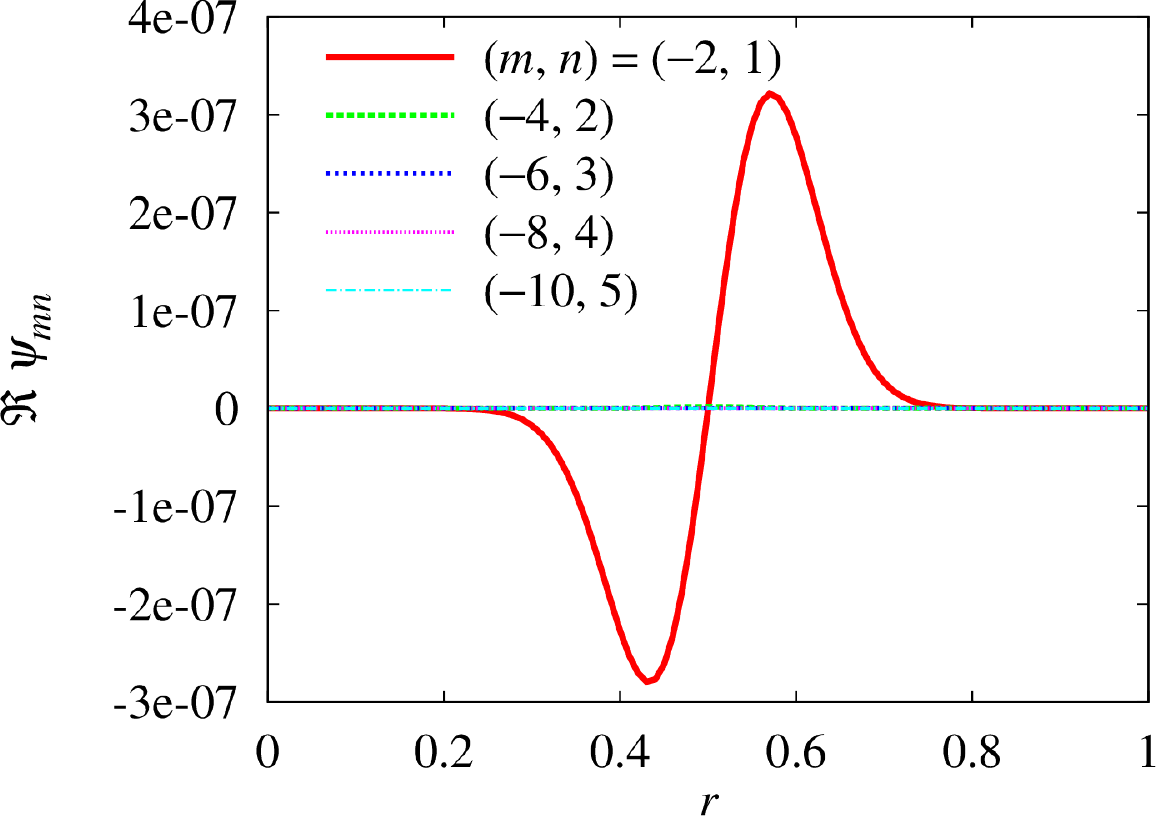}
 \subcaption{
 Radial profile of $\Re\, \psi_{mn}$.  The imaginary part is zero.
 }
 \label{fig:r-psi_r-t000000_0000-ini2}
\end{minipage} 
 \hfill
\begin{minipage}[t]{0.45\textwidth}
 \centering
 \includegraphics[width=\textwidth]{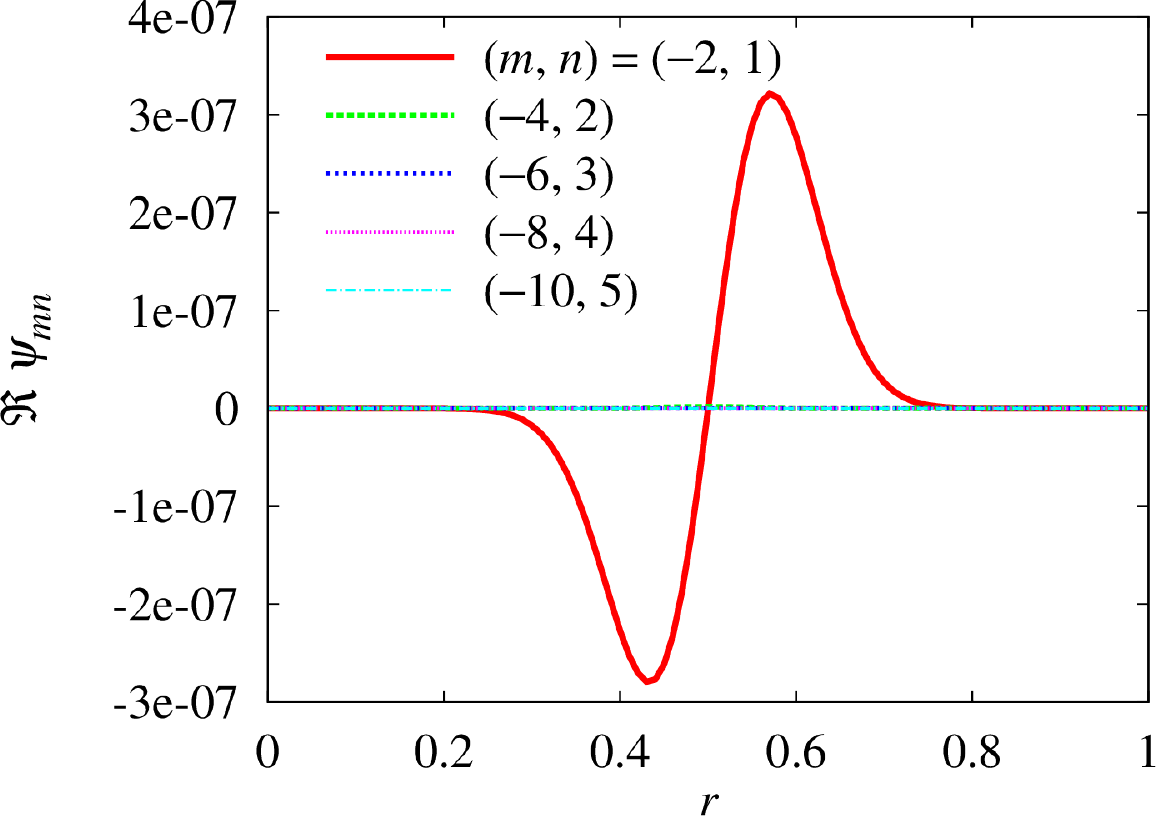}
 \subcaption{
 Radial profile of $\Re\, J_{mn}$.  The imaginary part is zero.
 }
 \label{fig:r-J_r-t000000_0000-ini2}
\end{minipage} 
 \caption{
 Radial profiles of the perturbed state that is used as an initial condition for  SA.
 The $(m, n) = (-2, 1)$ components are dominant, but they still have small amplitudes.
 }
 \label{fig:r-u-t000000_0000-ini2}
\end{figure*}

The SA results shown in the previous subsection \ref{subsubsec:acceleration}
were performed for the initial perturbation that has a larger
perturbed magnetic energy (${\cal O}(10^{-7})$) than the perturbed
kinetic energy (${\cal O}(10^{-12})$). 
In order to examine another case where the perturbed kinetic energy is
larger than the perturbed magnetic energy,
we chose a smaller $\tilde{\vphi}$ in Eq.~(\ref{eq:tvphi-perturb})
and a larger $\tilde{J}$ in Eq.~(\ref{eq:tJ-perturb}) for 
dynamically accessible perturbations.
The total energy of the system increased by the advection.
The radial profiles of the initial condition for SA are plotted in
Fig.~\ref{fig:r-u-t000000_0000-ini2}. 
The velocity part has about 200 times larger amplitudes and the magnetic part
has about 10 times smaller amplitudes compared with 
Fig.~\ref{fig:r-u-t000000_0000}.
The perturbed kinetic energy is ${\cal O}(10^{-7})$, which is larger than
the perturbed magnetic energy of ${\cal O}(10^{-9})$.
They are still in a linear regime.

\begin{figure}[h]
 \centering
 \includegraphics[width=0.5\textwidth]{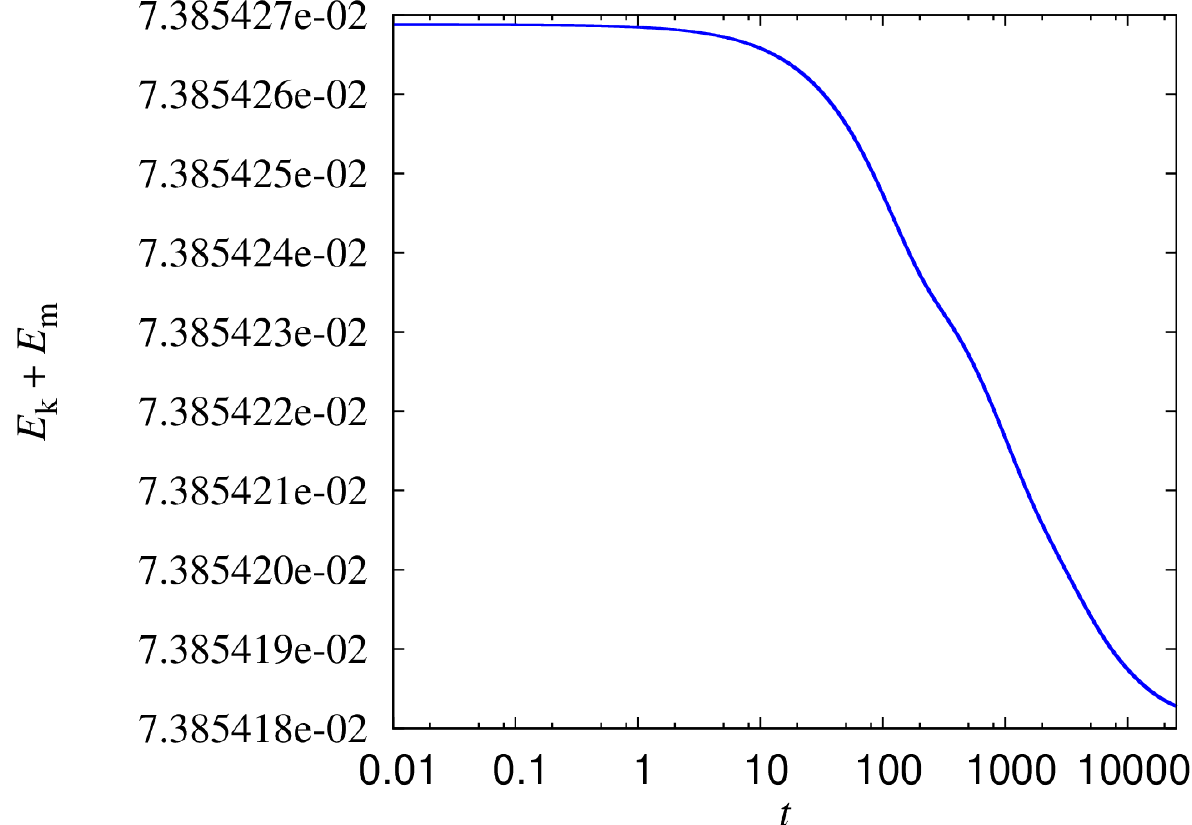}
 \caption{
 The total energy decreased in time by SA.  The perturbation amplitudes 
 become smaller in time as shown in Fig.~\ref{fig:r-u-t-mm2n1-ini2}
 }
 \label{fig:t-E-ini2}
\end{figure}

Starting from the initial state shown in
Fig.~\ref{fig:r-u-t000000_0000-ini2}, SA was performed.
For the acceleration, $F_{\rm max} = 10^{-3}$ and $\alpha_{\rm max} =
10^{7}$ were used.  The total energy decreased by SA as shown in
Fig.~\ref{fig:t-E-ini2}.
The simulation was stopped since the Fourier modes of the right-hand
sides of Eqs.~(\ref{eq:SA-vorticity-equation}) and
(\ref{eq:SA-Ohm-law}) for SA 
as well as the original low-beta reduced MHD
Eqs.~(\ref{eq:vorticity-equation}) and (\ref{eq:Ohm-law}) became smaller
than a threshold $10^{-8}$, although the energy looks still to be
decreasing.  Note that the horizontal axis is a log scale, thus the
decrease of energy is not so rapid.

\begin{figure*}
\begin{minipage}[t]{0.45\textwidth}
 \centering
 \includegraphics[width=\textwidth]{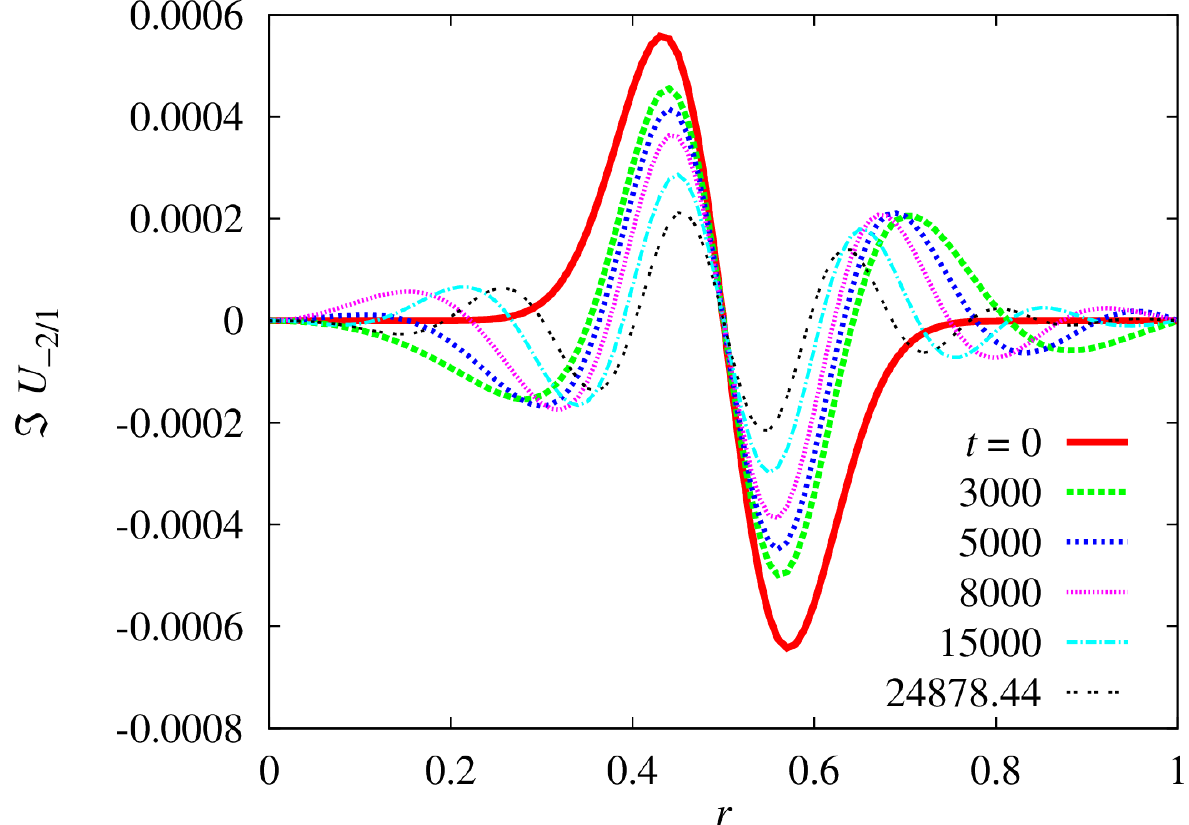}
 \subcaption{
 Radial profile of $\Im\, U_{-2,1}$.
 }
 \label{fig:r-U_i-t-mm2n1-ini2}
\end{minipage} 
 \hfill
\begin{minipage}[t]{0.45\textwidth}
 \centering
 \includegraphics[width=\textwidth]{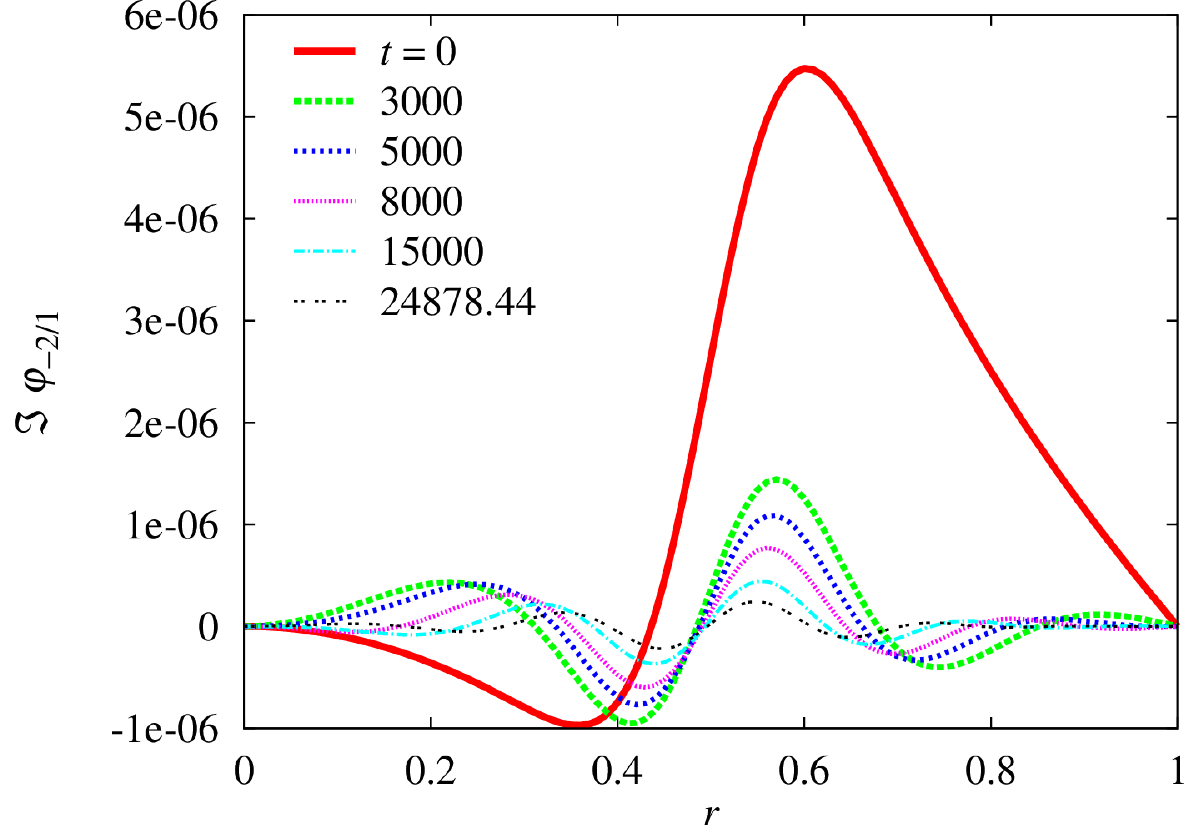}
 \subcaption{
 Radial profile of $\Im\, \vphi_{-2,1}$.
 }
 \label{fig:r-phi_i-t-mm2n1-ini2}
\end{minipage} 
\begin{minipage}[t]{0.45\textwidth}
 \centering
 \includegraphics[width=\textwidth]{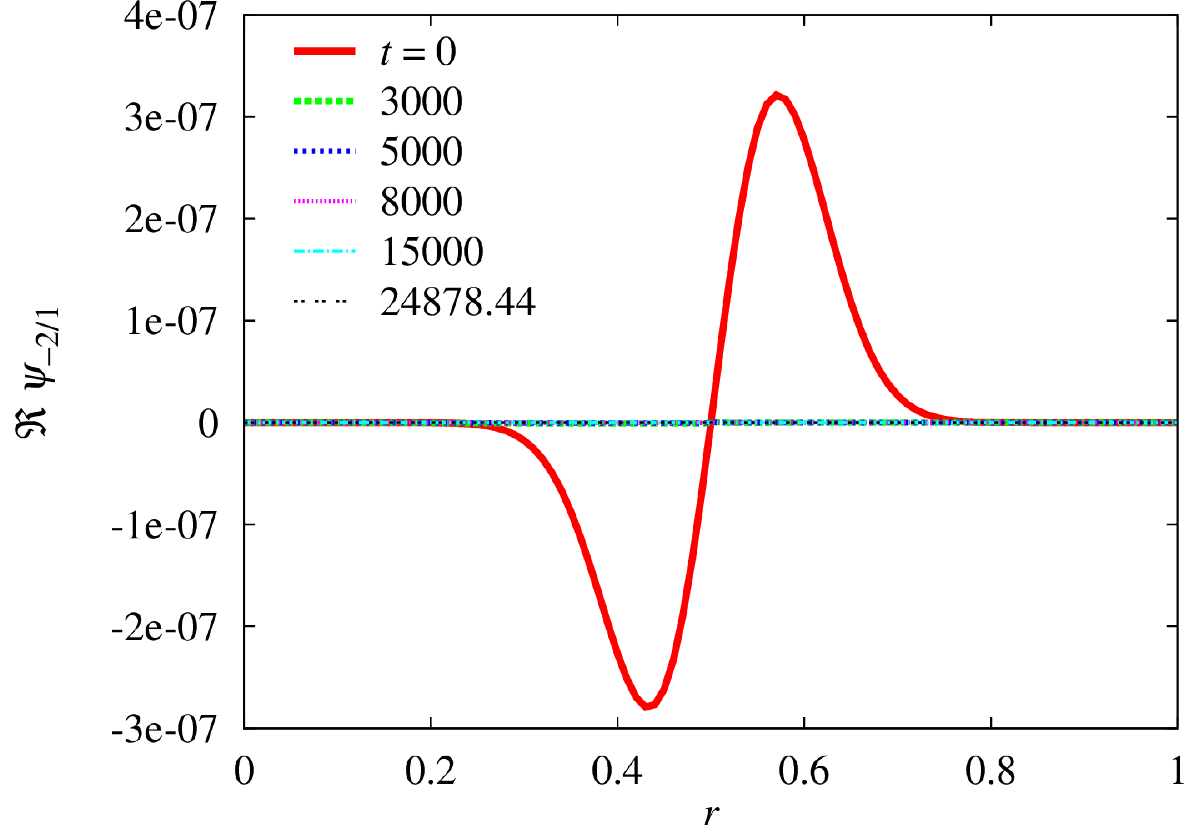}
 \subcaption{
 Radial profile of $\Re\, \psi_{-2,1}$.
 }
 \label{fig:r-psi_r-t-mm2n1-ini2}
\end{minipage} 
 \hfill
\begin{minipage}[t]{0.45\textwidth}
 \centering
 \includegraphics[width=\textwidth]{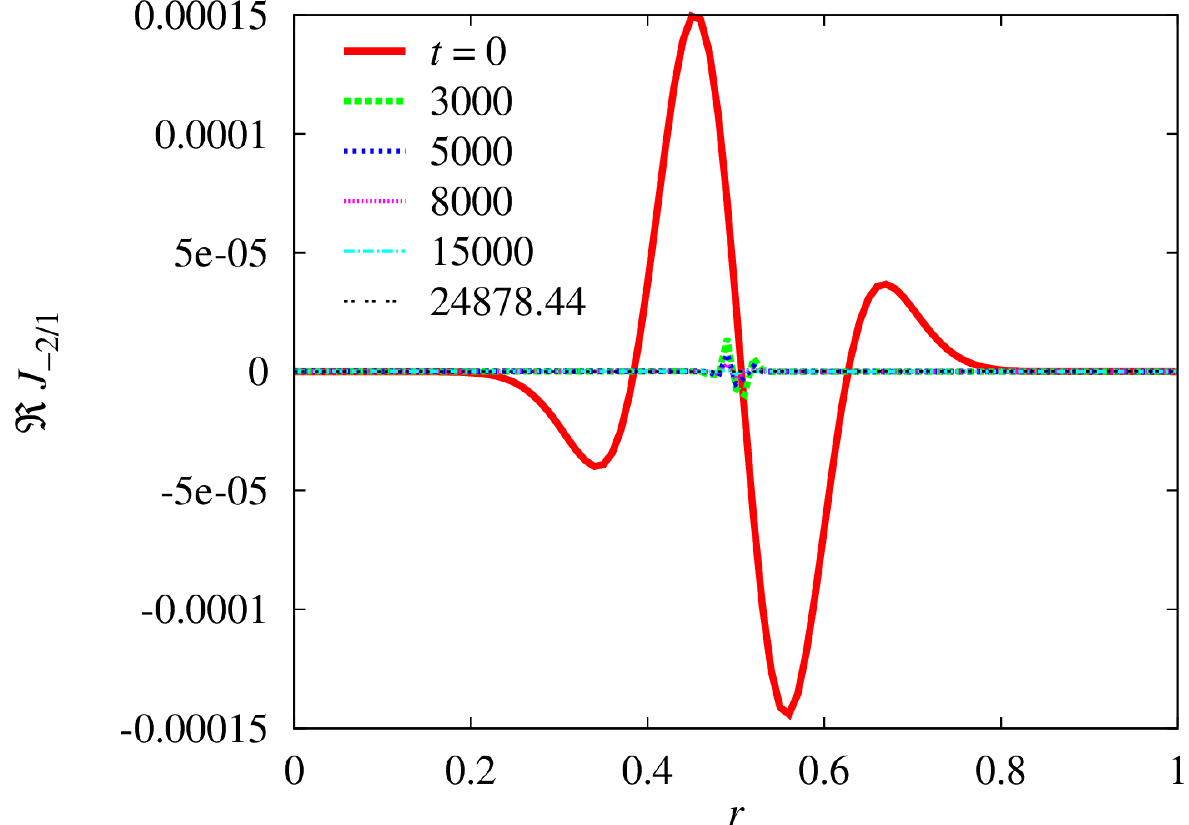}
 \subcaption{
 Radial profile of $\Re\, J_{-2,1}$.
 }
 \label{fig:r-J_r-t-mm2n1-ini2}
\end{minipage} 
 \caption{
 Radial profiles of the $(m, n) = (-2, 1)$ components are plotted  at
 several times during SA evolution.
The perturbation amplitudes decreased in time.
 }
 \label{fig:r-u-t-mm2n1-ini2}
\end{figure*}

The perturbation amplitudes decreased in time as shown in
Fig.~\ref{fig:r-u-t-mm2n1-ini2}.
Since the initial magnetic perturbation was small in this case,
the amplitudes of the magnetic part quickly became negligibly small.
The velocity part is still changing, decreasing the kinetic energy,
although it is already small.  In fact, not shown in the presented
figures, the magnetic energy became almost the same value as the
cylindrically symmetric state, while the kinetic energy of ${\cal
O}(10^{-9})$ still remained when the simulation was stopped.  However,
we observe that the amplitudes of the velocity part were disappearing.
The decrease of total energy with the disappearance of perturbation
again indicate that the equilibrium is stable.

\subsection{A case of unstable equilibrium}
\label{subsec:unstableCase}

\begin{figure}
 \centering
 \includegraphics[width=0.45\textwidth]{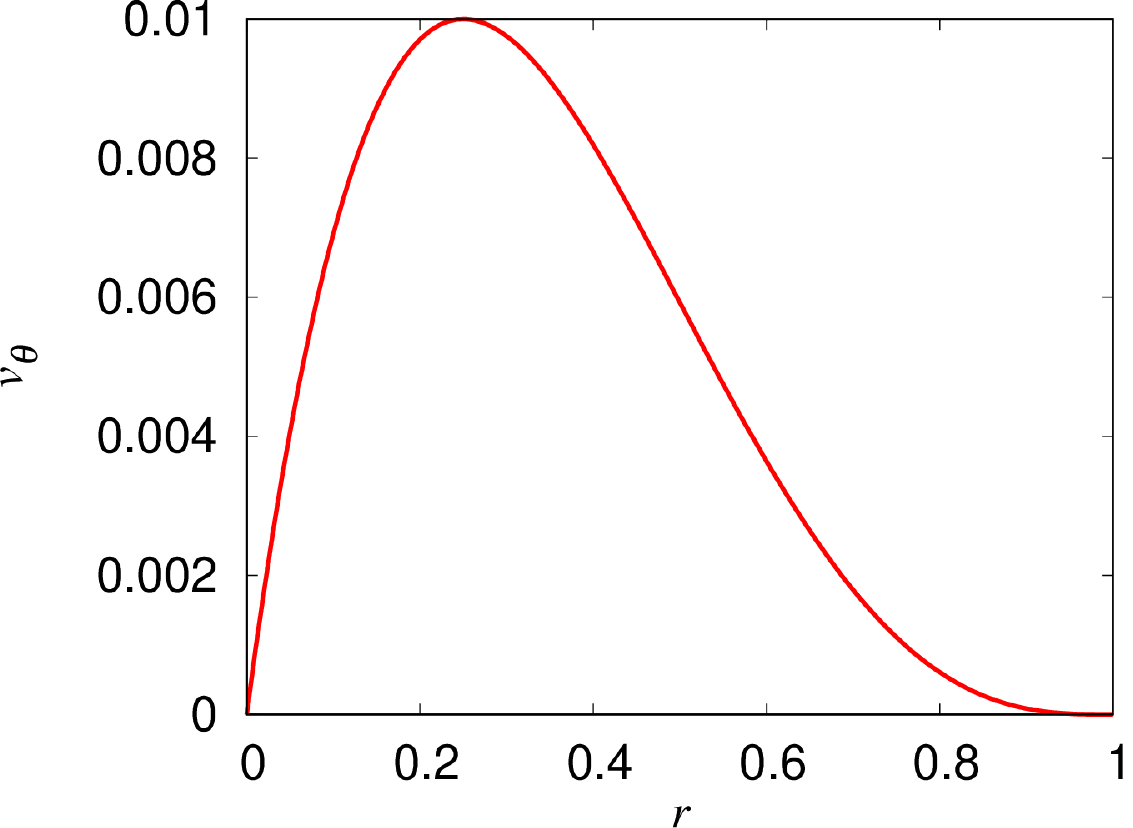}
 \caption{Poloidal rotation velocity $v_{\theta}$ profile. }
 \label{fig:r-vth}
\end{figure}

In this subsection, SA is performed for linear stability analysis of
an unstable equilibrium.
The safety factor profile is the same as Fig.~\ref{fig:r-q}.
The poloidal rotation profile is given by
\begin{equation}
 v_{\theta}(r)
  =
   \frac{v_{\theta {\rm max}} (\alpha + 1)^{\alpha + 1}}{\alpha^{\alpha}}
  r ( 1 - r )^{\alpha},
\end{equation}
where $\alpha$ is a positive parameter.  For the following numerical
results, $v_{\theta {\rm max}} = 0.01$ and $\alpha = 3$ were used.
The profile is plotted in Fig.~\ref{fig:r-vth}.

\begin{figure*}
 \centering
 \includegraphics[width=0.45\textwidth]{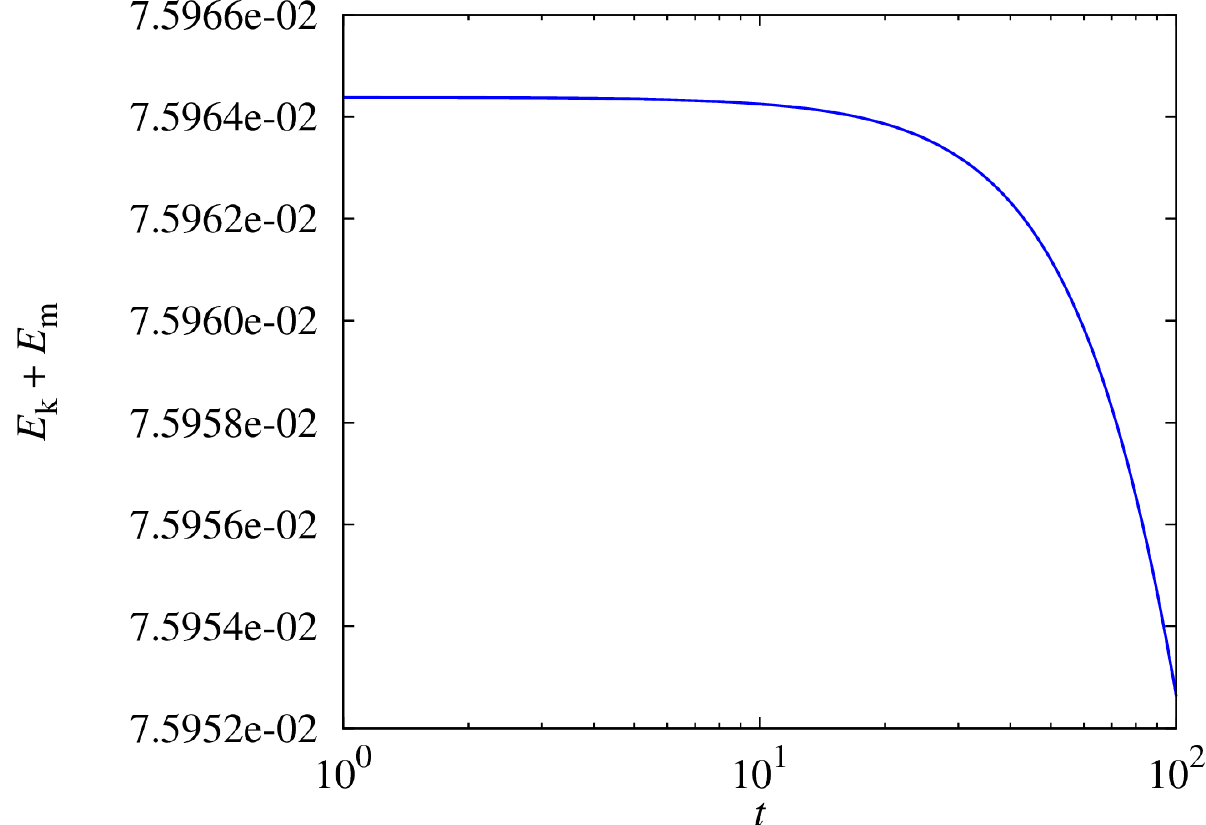}
 \caption{The total energy decreased in time by the dynamically
 accessible perturbation.}
 \label{fig:t-E-perturb-unstable}
\end{figure*}

This equilibrium was perturbed by the same advection fields 
of Eqs.~(\ref{eq:tvphi-perturb}) and (\ref{eq:tJ-perturb}) with the
amplitudes shown in Fig.~\ref{fig:r-tvphi-tJ-perturb}.
The total energy of the system decreased by the dynamically accessible
perturbation as shown in
Fig.~\ref{fig:t-E-perturb-unstable}.
This already means that there exists a state with a
lower energy than the cylindrically symmetric equilibrium.

\begin{figure*}
\begin{minipage}[t]{0.45\textwidth}
 \centering
 \includegraphics[width=\textwidth]{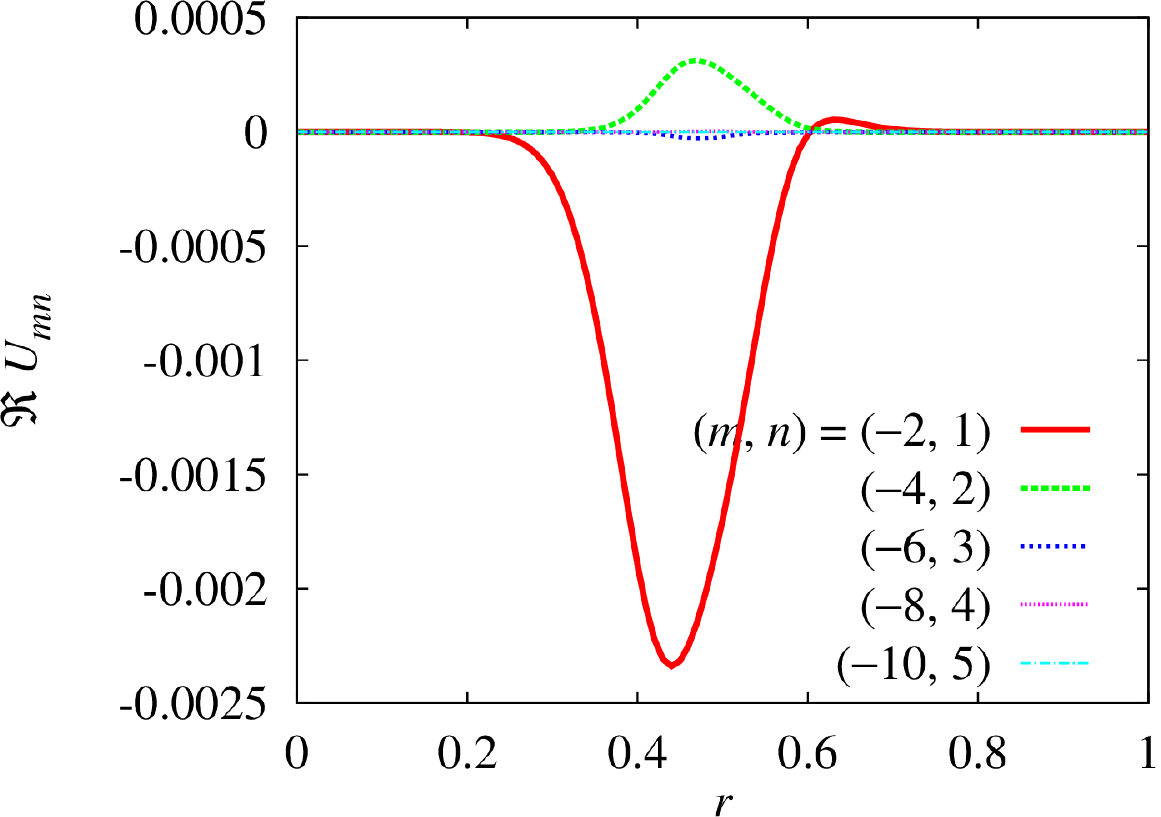}
 \subcaption{
 Radial profile of $\Re\, U_{mn}$.
 }
 \label{fig:r-U_r-t000000_0000-unstable}
\end{minipage} 
 \hfill
\begin{minipage}[t]{0.45\textwidth}
 \centering
 \includegraphics[width=\textwidth]{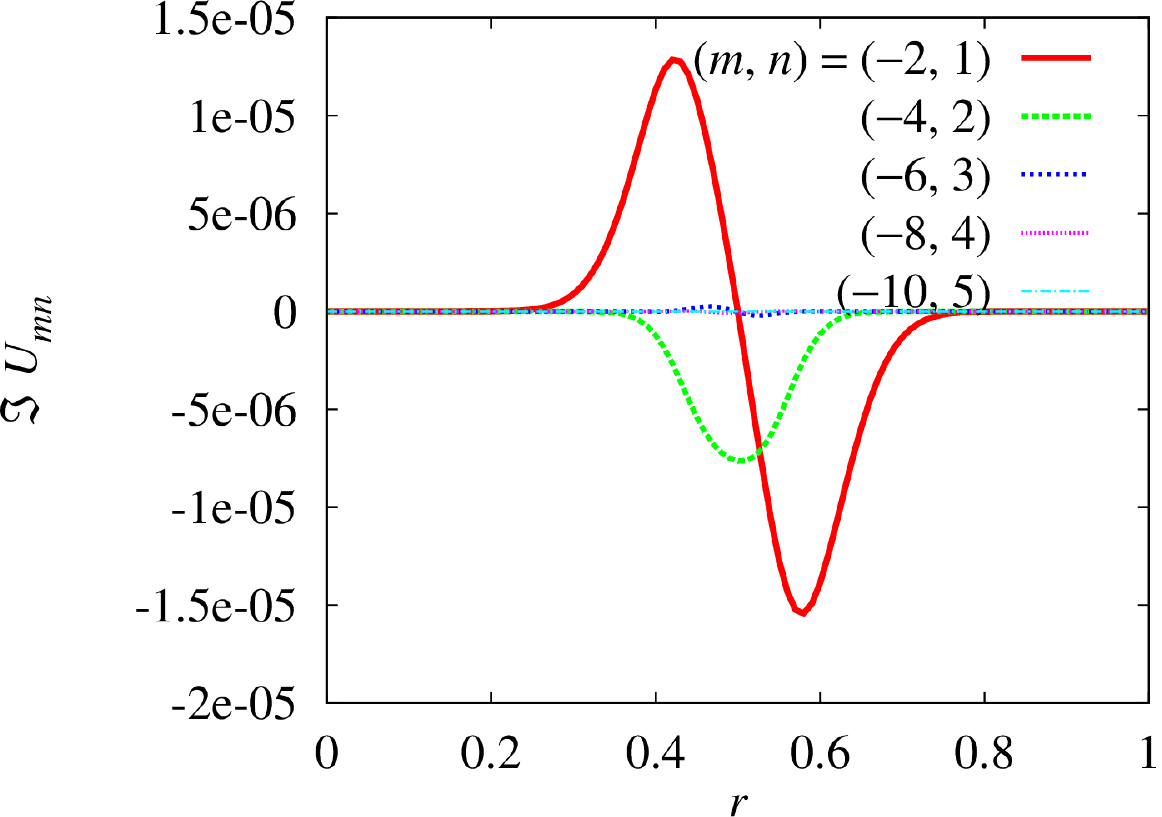}
 \subcaption{
 Radial profile of $\Im\, U_{mn}$.
 }
 \label{fig:r-U_i-t000000_0000-unstable}
\end{minipage} 
\begin{minipage}[t]{0.45\textwidth}
 \centering
 \includegraphics[width=\textwidth]{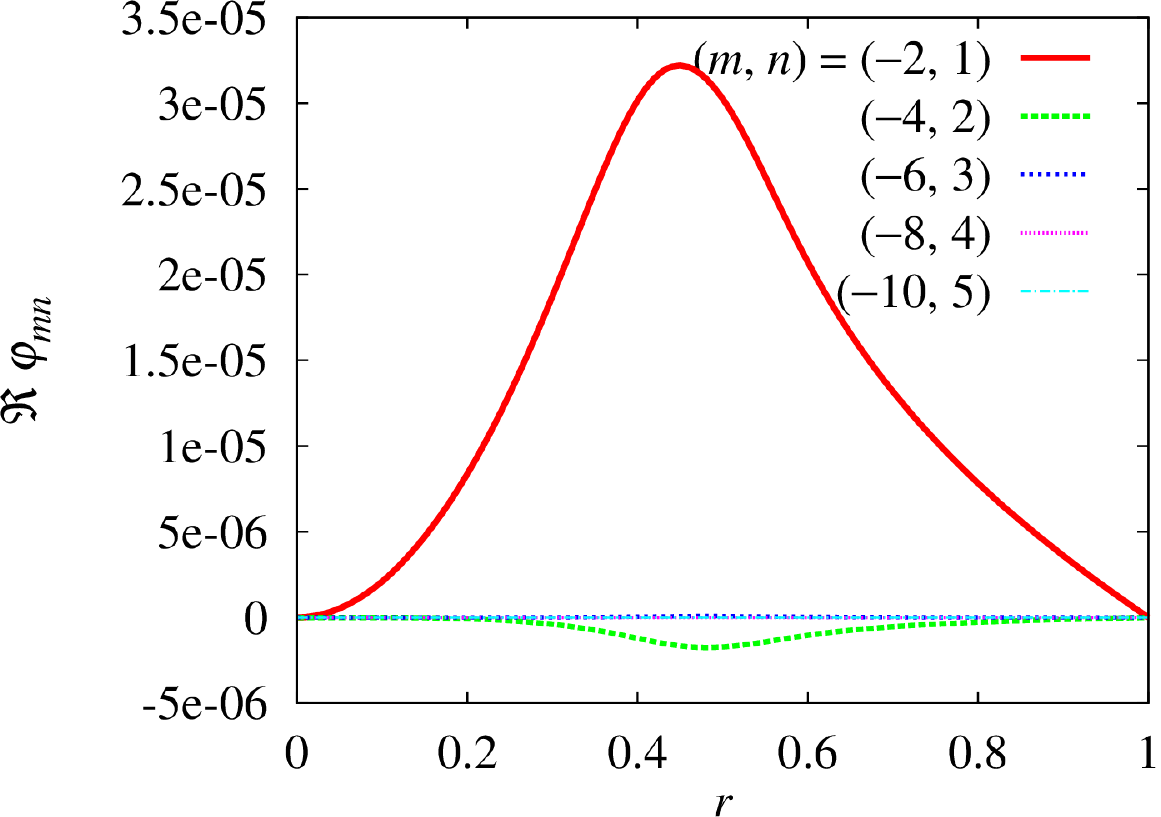}
 \subcaption{
 Radial profile of $\Re\, \vphi_{mn}$.
 }
 \label{fig:r-phi_r-t000000_0000-unstable}
\end{minipage} 
 \hfill
\begin{minipage}[t]{0.45\textwidth}
 \centering
 \includegraphics[width=\textwidth]{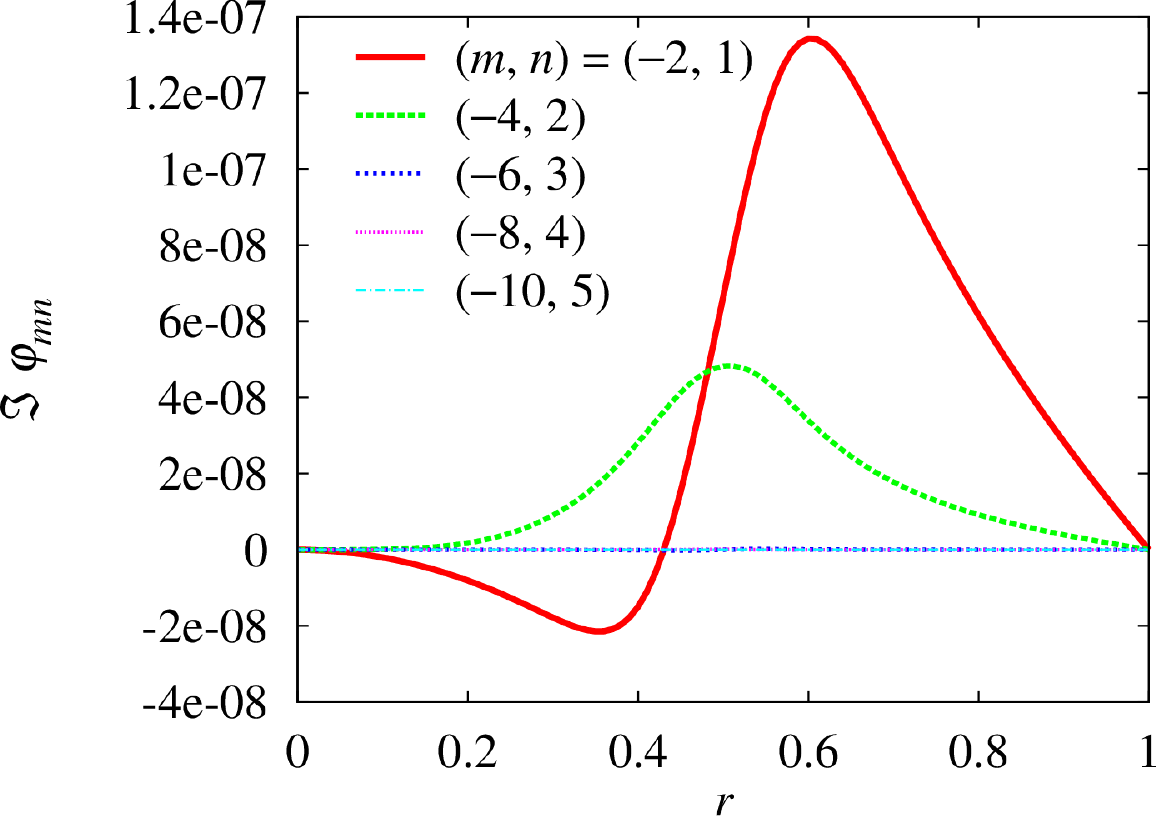}
 \subcaption{
 Radial profile of $\Im\, \vphi_{mn}$.
 }
 \label{fig:r-phi_i-t000000_0000-unstable}
\end{minipage} 
\begin{minipage}[t]{0.45\textwidth}
 \centering
 \includegraphics[width=\textwidth]{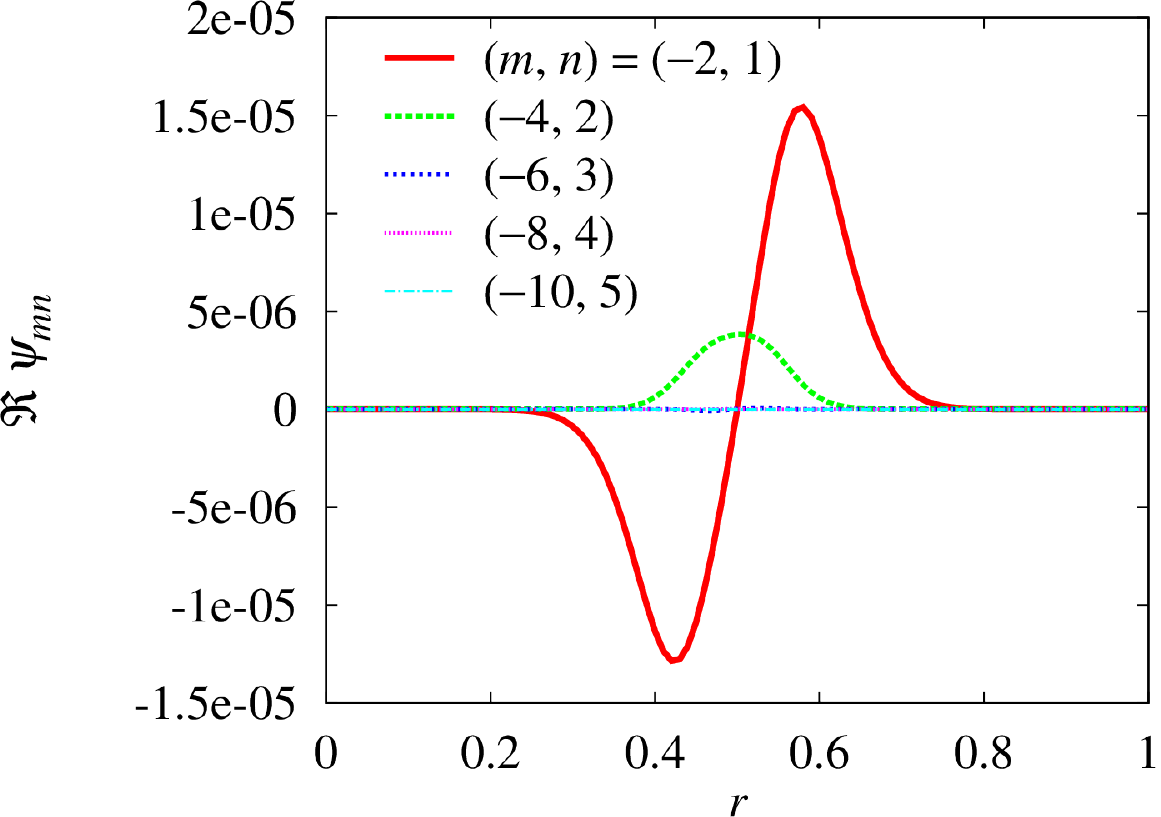}
 \subcaption{
 Radial profile of $\Re\, \psi_{mn}$.  The imaginary part is zero.
 }
 \label{fig:r-psi_r-t000000_0000-unstable}
\end{minipage} 
 \hfill
\begin{minipage}[t]{0.45\textwidth}
 \centering
 \includegraphics[width=\textwidth]{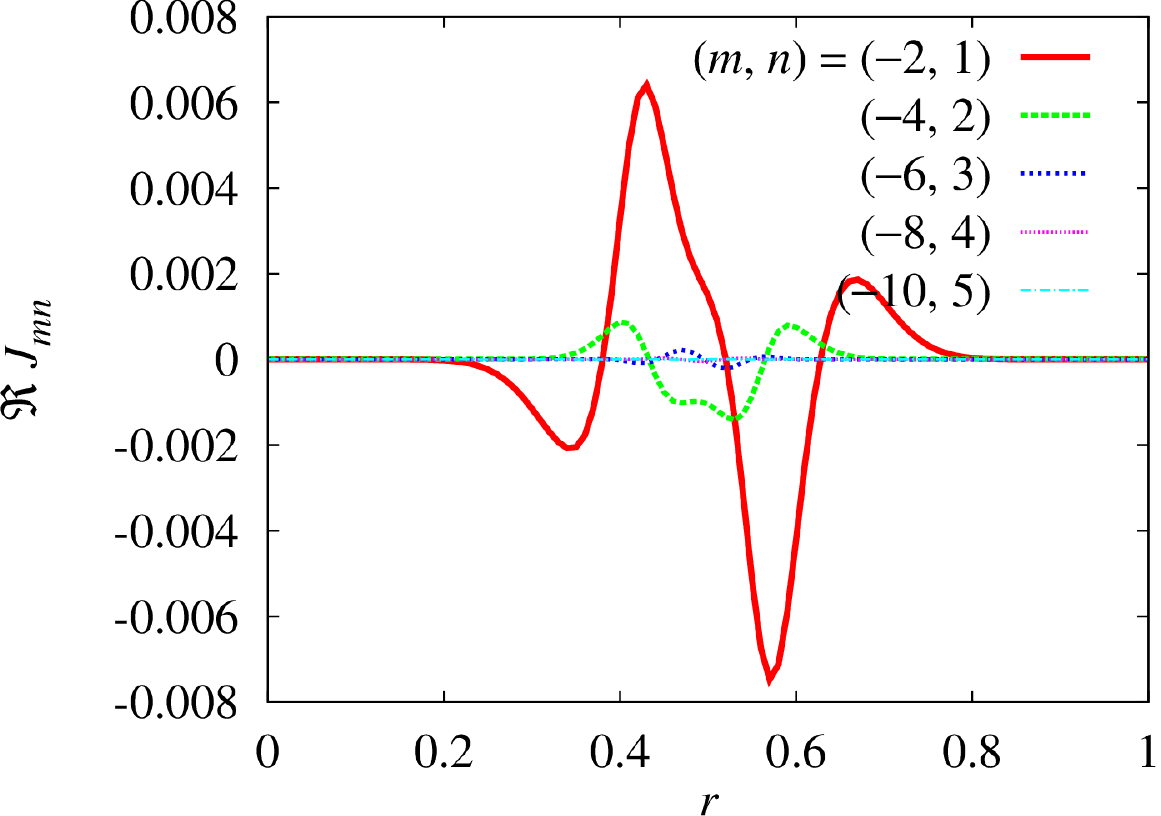}
 \subcaption{
 Radial profile of $\Re\, J_{mn}$.  The imaginary part is zero.
 }
 \label{fig:r-J_r-t000000_0000-unstable}
\end{minipage} 
 \caption{
 Radial profiles of the perturbed state that is used as an initial condition for  SA.
 The $(m, n) = (-2, 1)$ components are dominant, but they still have small amplitudes.
 }
 \label{fig:r-u-t000000_0000-unstable}
\end{figure*}

From a perturbed state with $t = 10$ in Fig.~\ref{fig:t-E-perturb-unstable},
we performed SA.  The radial profiles of the initial condition for 
SA is shown in Fig.~\ref{fig:r-u-t000000_0000-unstable}.
The $(m, n) = (-2, 1)$ components are dominant, although
the $(m, n) = (-4, 2)$ and $(m, n) = (-6, 3)$ components are still
visible.  The amplitudes are small and in a linear regime.

\begin{figure}[h]
 \centering
 \includegraphics[width=0.5\textwidth]{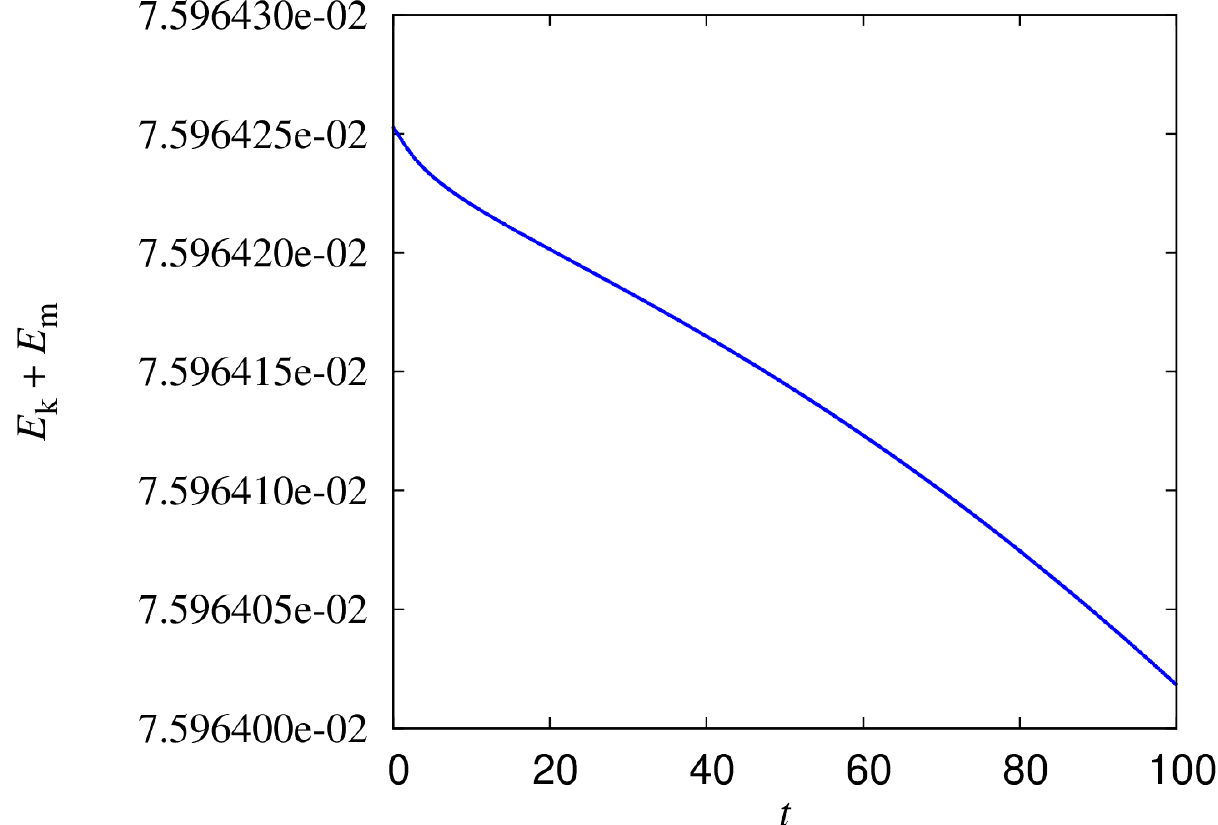}
 \caption{
 The total energy decreased in time by SA, although the perturbation
 amplitudes grew in time as shown in Fig.~\ref{fig:r-u-t-mm2n1-unstable}
 }
 \label{fig:t-E-unstable}
\end{figure}

The time evolution of the total energy by SA is plotted in
Fig.~\ref{fig:t-E-unstable}.
The total energy decreased by SA.
For the kernel of the double bracket, $F_{\rm max} = 10^{-3}$ and
$\alpha_{\rm max} = 10^{7}$ were used.
After $t \simeq 100$, the radial profiles of perturbed quantities started to
oscillate in $r$, and the conservation of magnetic helicity was
apparently violated.
Therefore the time evolution is plotted only for $t \leq 100$.
Linear eigenmode analysis without dissipation for this equilibrium
gave us multiple unstable modes, although the numerical technique used
was very simple.  The eigenmodes look singular.
The physical situation seems to be numerically difficult,
and oscillation in $r$ occurred. 

\begin{figure*}
\begin{minipage}[t]{0.35\textwidth}
 \centering
 \includegraphics[width=\textwidth]{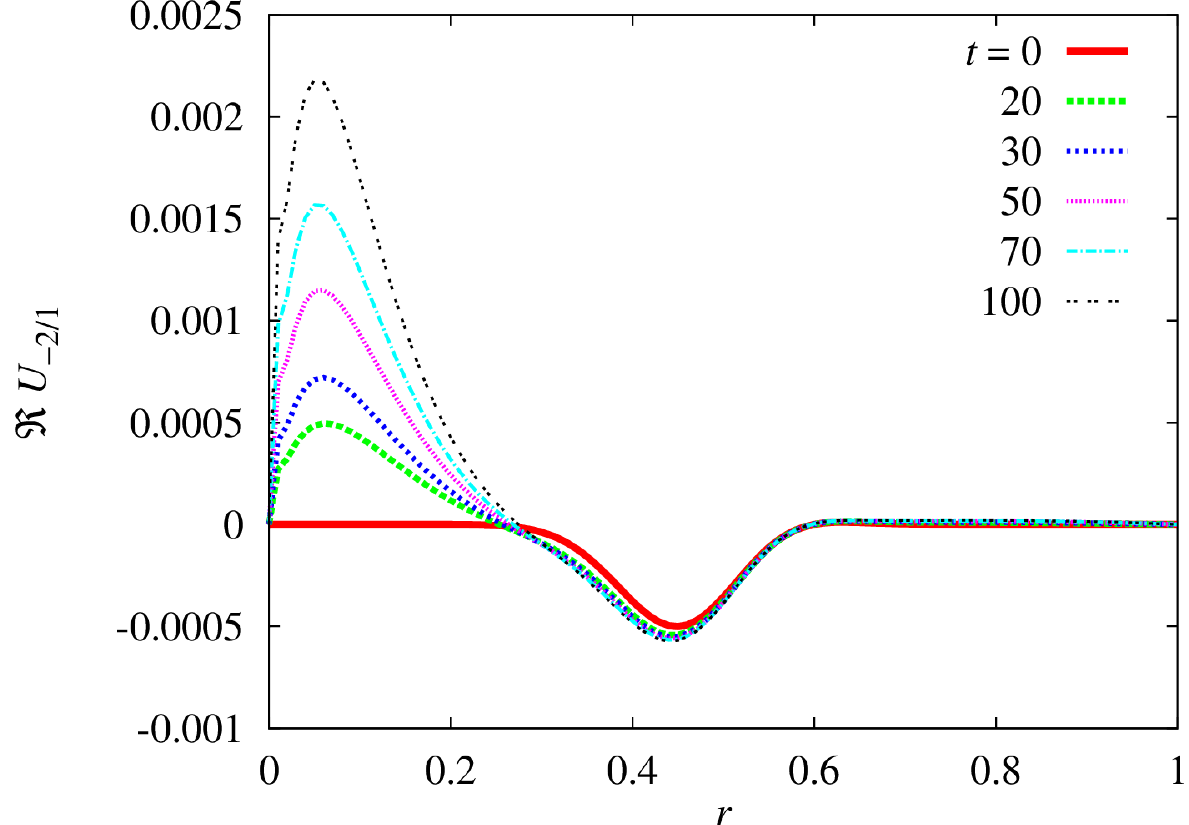}
 \subcaption{
 Radial profile of $\Re\, U_{-2,1}$.
 }
 \label{fig:r-U_r-t-mm2n1-unstable}
\end{minipage} 
 \hfill
\begin{minipage}[t]{0.35\textwidth}
 \centering
 \includegraphics[width=\textwidth]{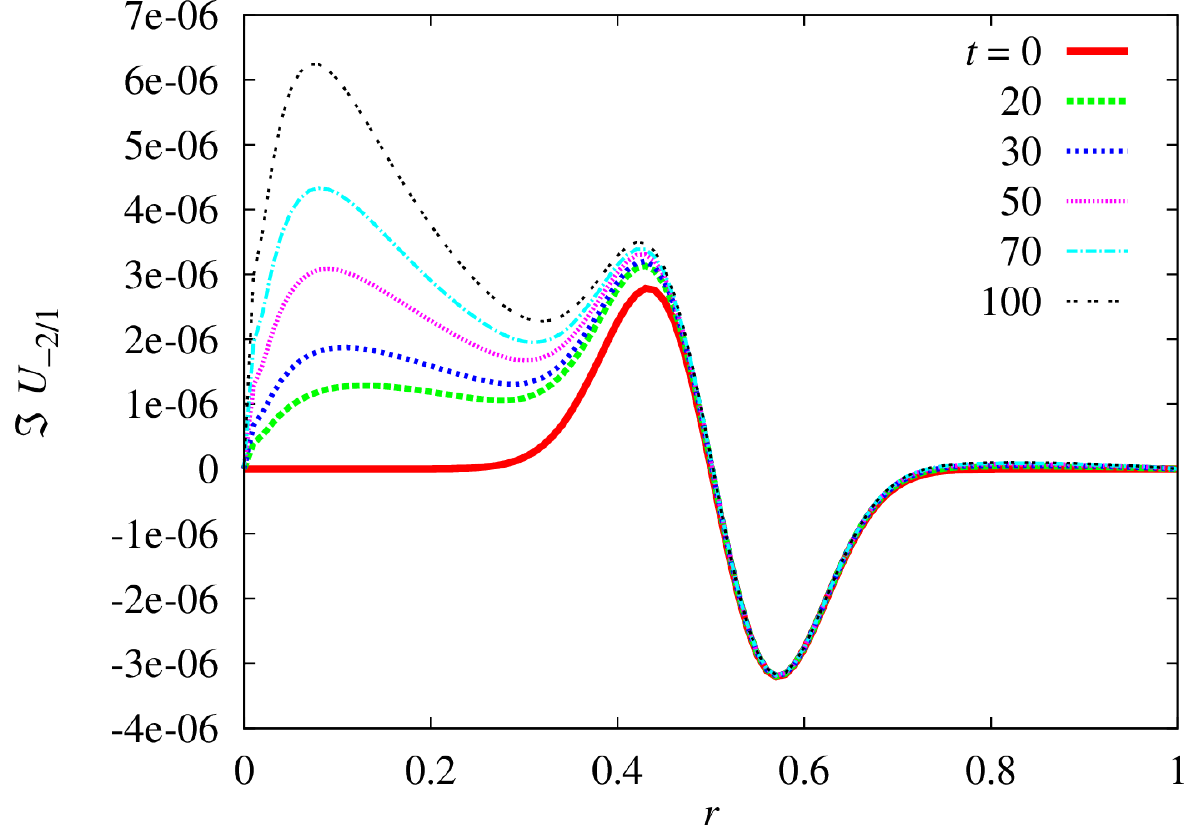}
 \subcaption{
 Radial profile of $\Im\, U_{-2,1}$.
 }
 \label{fig:r-U_i-t-mm2n1-unstable}
\end{minipage} 
\begin{minipage}[t]{0.35\textwidth}
 \centering
 \includegraphics[width=\textwidth]{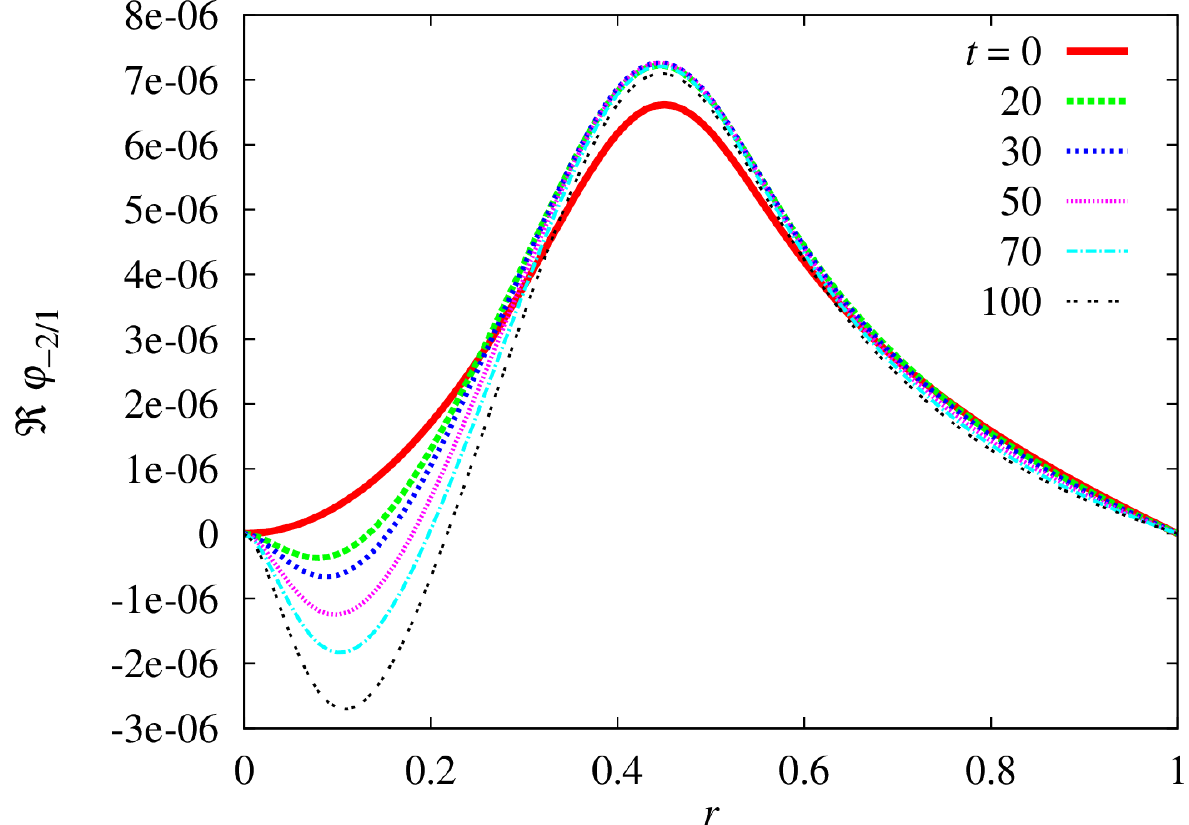}
 \subcaption{
 Radial profile of $\Re\, \vphi_{-2,1}$.
 }
 \label{fig:r-phi_r-t-mm2n1-unstable}
\end{minipage} 
 \hfill
\begin{minipage}[t]{0.35\textwidth}
 \centering
 \includegraphics[width=\textwidth]{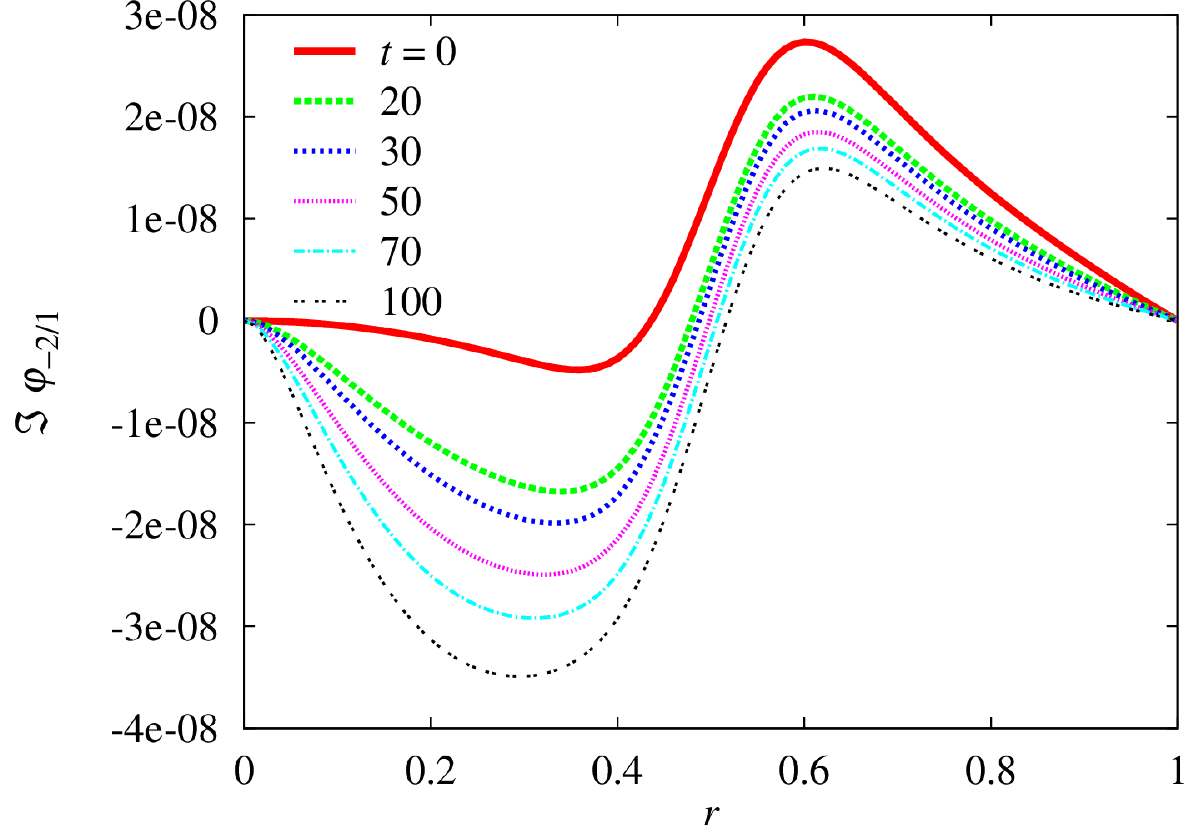}
 \subcaption{
 Radial profile of $\Im\, \vphi_{-2,1}$.
 }
 \label{fig:r-phi_i-t-mm2n1-unstable}
\end{minipage} 
\begin{minipage}[t]{0.35\textwidth}
 \centering
 \includegraphics[width=\textwidth]{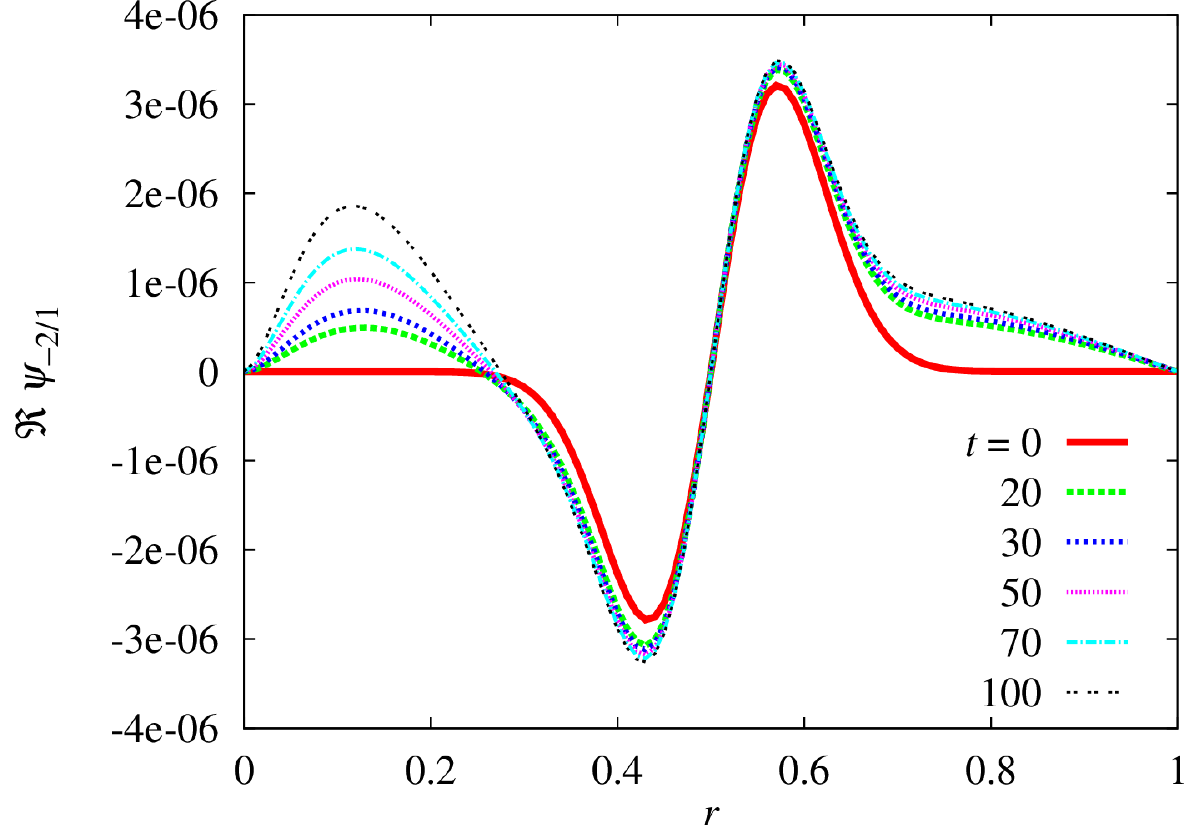}
 \subcaption{
 Radial profile of $\Re\, \psi_{-2,1}$.
 }
 \label{fig:r-psi_r-t-mm2n1-unstable}
\end{minipage} 
 \hfill
\begin{minipage}[t]{0.35\textwidth}
 \centering
 \includegraphics[width=\textwidth]{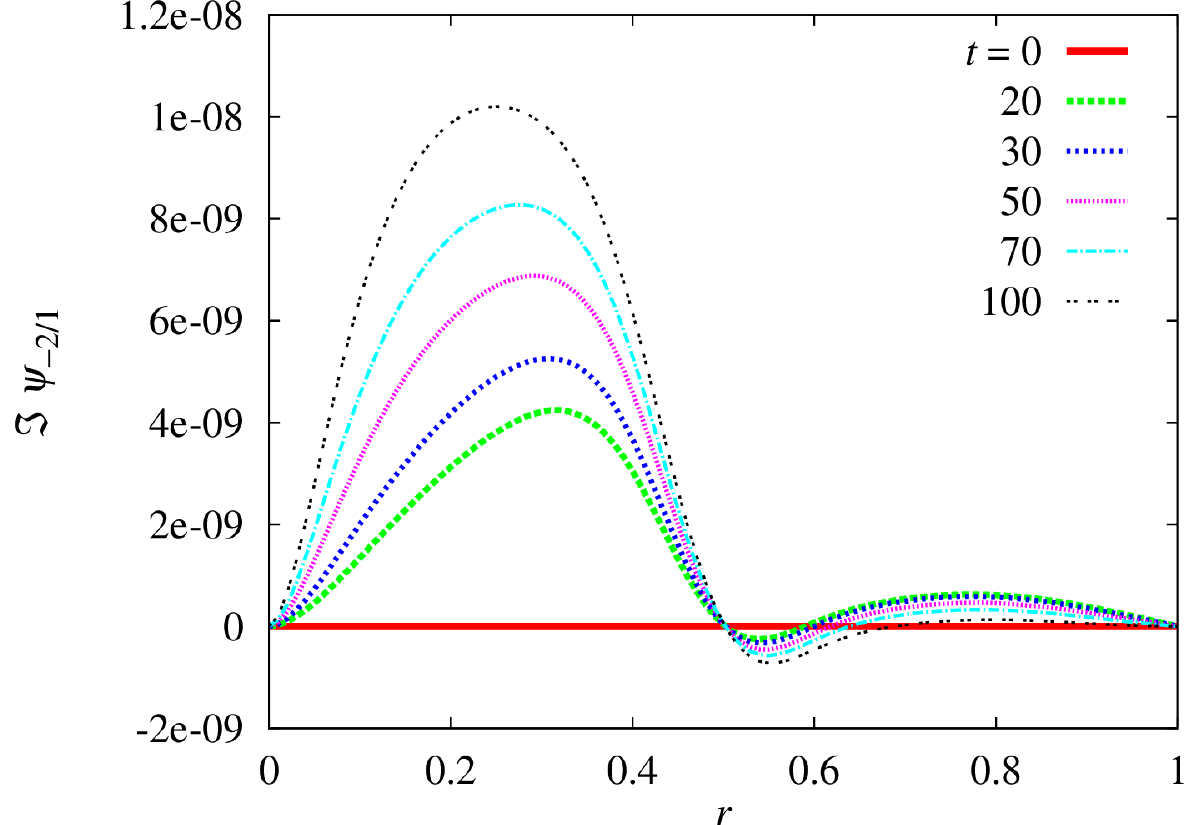}
 \subcaption{
 Radial profile of $\Im\, \psi_{-2,1}$.
 }
 \label{fig:r-psi_i-t-mm2n1-unstable}
\end{minipage} 
\begin{minipage}[t]{0.35\textwidth}
 \centering
 \includegraphics[width=\textwidth]{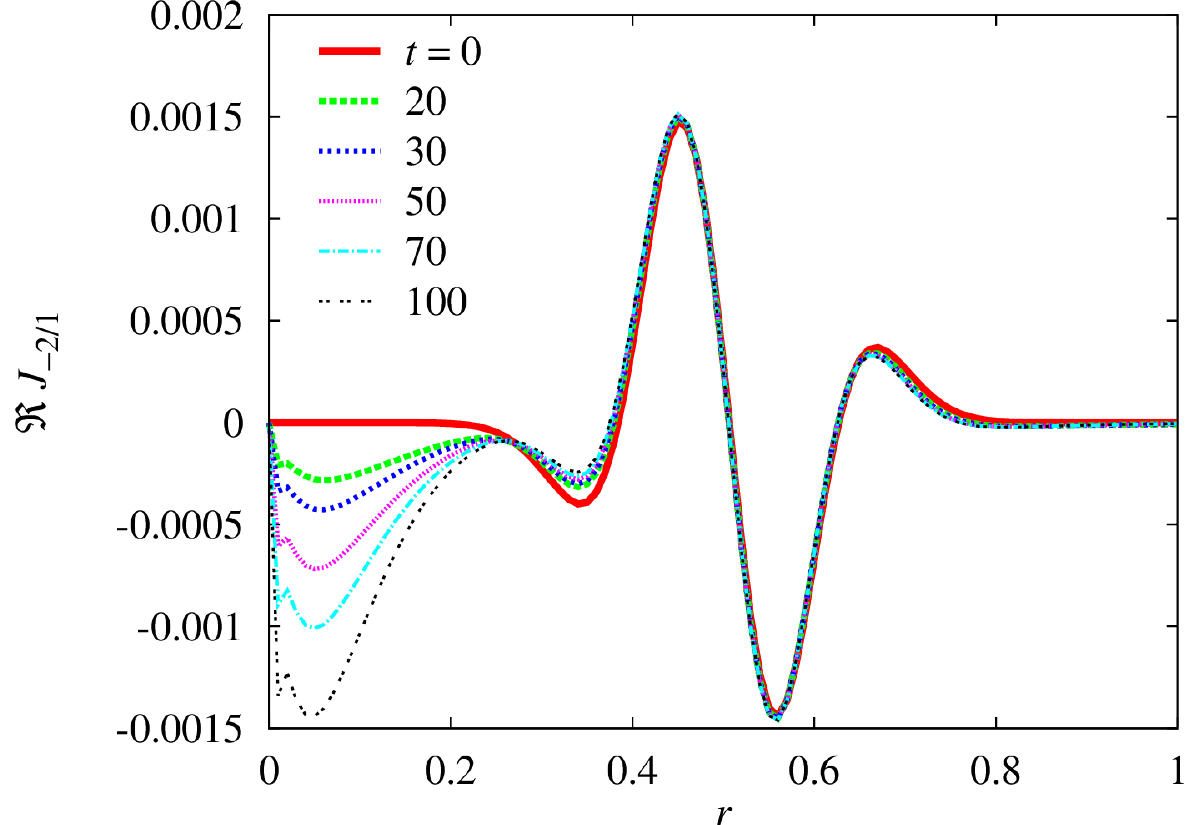}
 \subcaption{
 Radial profile of $\Re\, J_{-2,1}$.
 }
 \label{fig:r-J_r-t-mm2n1-unstable}
\end{minipage} 
 \hfill
\begin{minipage}[t]{0.35\textwidth}
 \centering
 \includegraphics[width=\textwidth]{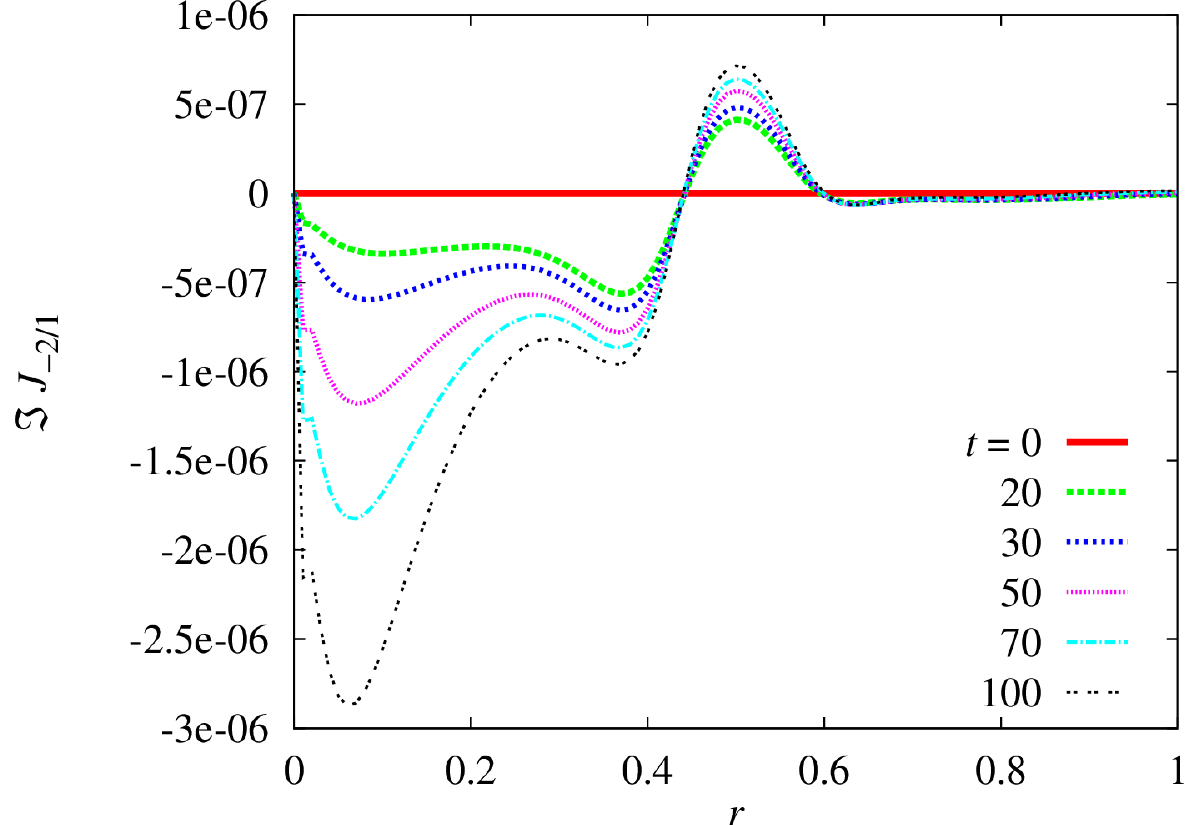}
 \subcaption{
 Radial profile of $\Im\, J_{-2,1}$.
 }
 \label{fig:r-J_i-t-mm2n1-unstable}
\end{minipage} 
 \caption{
 Radial profiles of the $(m, n) = (-2, 1)$ components are plotted  at
 several times during the SA evolution.
The perturbation amplitudes grew in time.
 }
 \label{fig:r-u-t-mm2n1-unstable}
\end{figure*}

Although the total energy decreased in time by SA,
the perturbation amplitudes grew in time as shown in
Fig.~\ref{fig:r-u-t-mm2n1-unstable}.
As noted above, the radial profiles started to oscillate after $t \simeq
100$, and thus another equilibrium without cylindrical symmetry was not
obtained.  However, the growth of the perturbation under decreasing
energy shows that the original equilibrium is at least linearly unstable.

\subsection{Stability of equilibrium via SA}
\label{subsec:stability}

Let us now consider how SA can be used to find a stable equilibrium.
Finding a stable equilibrium is rather time consuming since we have to
observe that the given perturbation actually disappears by SA.
We tried two kinds of dynamically accessible perturbations:
one with a larger perturbed magnetic energy, the other with
a larger perturbed kinetic energy, although both are still in a linear regime.
In particular, we consider the equilibrium  state obtained in
Sec.~\ref{subsubsec:acceleration}, which has almost zero kinetic energy, 
it being ${\cal O}(10^{-15})$.  Also the magnetic energy reaches the
stationary value and does not seem to further decrease. The magnetic
components with finite $m$ and/or $n$ disappear as time progresses  as
shown in Fig.~\ref{fig:r-u-t-mm2n1}, although only the $(m, n) = (-2,
1)$ components are shown.
We obtained similar results in Sec.~\ref{subsubsec:anotherInitialPerturbation}
for the other initial perturbation.
Therefore,  SA seems to recover the original cylindrical symmetric
equilibrium. This is of course reasonable since the original equilibrium
is linearly stable against ideal MHD modes.   In fact, the total energy
is almost the same as that of  the original equilibrium;  it is larger
than the original one by ${\cal O}(10^{-10})$ for the case of
Sec.~\ref{subsubsec:acceleration}
and ${\cal O}(10^{-9})$ for the case of 
Sec.~\ref{subsubsec:anotherInitialPerturbation}, of which relative
magnitudes to the total energy of the original equilibrium are
${\cal O}(10^{-9})$ and ${\cal O}(10^{-8})$, respectively.

Thus, here we have shown two cases where  SA recovers the original
equilibrium if it is ideal MHD stable. Another case was performed with
$r_{0} = 0.8$ in Eqs.~(\ref{eq:tvphi-perturb}) and
(\ref{eq:tJ-perturb}), 
with similar amplitudes as in Fig.~\ref{fig:r-tvphi-tJ-perturb}.
The perturbation in this case exists mainly in the region outside the  $q=2$ surface, which is resonant with the family of $(m, n) = (-2, 1)$ harmonics.  For this case we obtained  qualitatively similar results.

For an unstable equilibrium case, SA finds growth of the dynamically
accessible perturbation in time under decreasing total energy.
We expect the SA procedure, in principle, will converge to another
equilibrium if there is one on the same Casimir leaf and the
perturbation lies within its basin of attraction.  
In the case tried in this paper, however, the physical situation seems
to be tough for numerical study, and only linear stability was examined.

\section{Conclusions}
\label{sec:conclusion}

In this paper, simulated annealing (SA) was applied to low-beta reduced
magnetohydrodynamics (MHD) in cylindrical geometry.  Specifically, SA
was used to verify the stability of a cylindrically symmetric
equilibrium,  which is known to be linearly stable or unstable against
ideal MHD modes.  To this end, a dynamically accessible perturbation was
added to the cylindrically symmetric equilibrium by introducing
advection fields for perturbing the equilibrium without leaving the
Casimir leaf on which the original symmetric equilibrium exists.  The
dynamically accessible perturbation increased (decreased) the total
energy of the system if the original equilibrium is linearly stable
(unstable). Then SA was performed to monotonically decrease the total
energy. It was found that the SA reasonably recovers the original,
cylindrically symmetric equilibrium when it is stable.
On the other hand, the perturbation grew when the original equilibrium
is unstable.
Since low-beta reduced MHD is a two-field model,  it has two advection fields for relaxation.  It was found that for practical implementation of SA,  it is  important to balance the order of magnitude of the two advection fields for SA relaxation to occur on a short time scale.  To achieve such efficient relaxation,  time dependence was introduced in the symmetric kernel of SA, in such a way as to  balance the orders of magnitude of the two advection fields, thereby leading to accelerated convergence to the minimum energy state. 
The acceleration is particularly necessary for a stable equilibrium,
since we have to observe that the perturbation actually disappears.

In essence, the SA method provides a general prescription for building
from a Hamiltonian system an alternative non-Hamiltonian system with
asymptotic stability to an equilibrium state, with the stability
provided by an energy function serving as a Lyapunov
function. Convergence to the equilibrium requires that the Lyapunov
function have a local extremum, which is enough to show stability for
the original Hamiltonian system. Systems that are spectrally stable need
not have Lyapunov functions (because of negative energy modes, see e.g. Ref.~\onlinecite{Morrison-1998}) so the SA method here provides a method that singles out systems that are likely to be  nonlinearly stable.  Given this general framework, there are many avenues for further explorations in the context of fluid and kinetic plasma models.

\begin{acknowledgments}
MF was supported by JSPS KAKENHI Grant Number JP21K03507 while PJM was supported by the U.S. Dept. of Energy Contract \# DE-FG05-80ET-53088 and the Humboldt  Foundation.  We both would like to acknowledge the JIFT program for support of MF to visit IFS in the Spring of 2019 when a portion of this work was carried out. 
\end{acknowledgments}


%

\end{document}